\documentclass[journal]{IEEEtran}
\usepackage[cmex10]{amsmath}
\usepackage{cite}
\usepackage{url}
\usepackage{color}

\usepackage{amssymb}
\usepackage{path}
\usepackage{verbatim}
\usepackage{algorithm}
\usepackage{framed}
\usepackage{algpseudocode}

\usepackage{color}
\usepackage{amsmath}
\usepackage{cases}
\usepackage{theorem}
\usepackage{subfig}
\usepackage{subfloat}

\usepackage{graphicx}

\usepackage{pgfplots}
\usepackage{pgfplotstable}
\pgfplotsset{compat=newest}

\newtheorem{theorem}{Theorem}
{ \theorembodyfont{\normalfont} 

}

\newcounter{enumctr}

\DeclareFontFamily{U}{mathx}{\hyphenchar\font45}
\DeclareFontShape{U}{mathx}{m}{n}{<-> mathx10}{}
\DeclareSymbolFont{mathx}{U}{mathx}{m}{n}
\DeclareMathAccent{\widebar}{0}{mathx}{"73}

\ifCLASSINFOpdf
\else
\fi
\begin{document}
%
\title{Smart Procurement of Naturally Generated Energy (SPONGE) for Plug-in Hybrid Electric Buses}

%
%

\author{Joe Naoum-Sawaya,~\IEEEmembership{Member,~IEEE,}
        Emanuele Crisostomi,~\IEEEmembership{Member,~IEEE,}
        Mingming Liu,~\IEEEmembership{Member,~IEEE,}
        Yingqi Gu, ~\IEEEmembership{Graduate~Student~Member, ~IEEE,}
        and~Robert~Shorten,~\IEEEmembership{Senior~Member,~IEEE}
	\thanks{J. N. Sawaya is with the Ivey Business School, University of Western Ontario, 1255 Western Rd, London, Canada (e-mail: jnaoum-sawaya@ivey.ca).}
	
	\thanks{E. Crisostomi is with the Department of Energy, Systems, Territory and Constructions Engineering, University of Pisa, Italy (e-mail: emanuele.crisostomi@dsea.unipi.it).}
	
	\thanks{Y. Gu and M. Liu and R. Shorten are with the School of Electrical, Electronic and Communications Engineering, University College Dublin, Ireland (e-mail: yingqi.gu@ucdconnect.ie; mingming.liu@ucd.ie).}
	
	\thanks{R. Shorten is also with IBM Research, Dublin, Ireland (e-mail: robshort@ie.ibm.com).}}

%
%

\markboth{Journal of \LaTeX\ Class Files,~Vol.~6, No.~1, January~2007}%
{Shell \MakeLowercase{\textit{et al.}}: Bare Demo of IEEEtran.cls for Journals}
%



\maketitle

\begin{abstract}
We discuss a recently introduced  ECO-driving concept known as SPONGE in the context of Plug-in Hybrid Electric Buses (PHEB)'s.Examples are given to illustrate the benefits of this approach to ECO-driving. Finally, distributed algorithms to realise SPONGE are discussed, paying attention to the privacy implications of the underlying optimisation problems.
\end{abstract}


Note to Practitioners:
\begin{abstract}
In this paper we present a new idea for ECO-driving for buses. It is an IoT concept - that instead of connecting devices in space, connects devices in time via forecasting engines. Basically, a bus uses knowledge of the available energy at the next charging step, to optimise its performance beforehand. The system can be implemented using available (free) forecasting engines, and existing distributed optimisation tools. A sample implementation is described using a Toyota plug-in 
Prius (as a proxy for a hybrid bus). Apart from the forecasting and optimisation analytics, the only additional work needed was the development of an interface unit to control EV mode of the vehicle, and the development of a smart-phone app. Future work will investigate impacts of our approach on the grid, integration of the ideas into the hybrid drive cycle, and using driver behaviour as an input into the design of the utility functions.
\end{abstract}

Primary and Secondary Keywords
\begin{IEEEkeywords}
Primary Topics: Number 7, Number 8; 
Secondary Topic Keywords: Routing, Distributed Systems 
\end{IEEEkeywords}

%
\IEEEpeerreviewmaketitle

\section{Introduction}
%
%
%
%
\IEEEPARstart{W}{e} discuss a recently introduced holistic ECO-driving concept known as SPONGE in the context of Plug-in Hybrid Electric Buses (PHEB)'s. PHEB's are increasingly seen as an effective tool in combating air pollution in our cities, and as a tool for reducing our cities reliance on fossil fuels (thereby reducing greenhouse gas emissions) \cite{PHEBs, Example_US}\footnote{See https://chargedevs.com/newswire/ultramodern-plug-in-buses-go-into-service-in-gothenburg/~ for further examples.}. Consequently, the design and operation of such buses has been the subject of much research interest.  Hitherto, significant research effort has focused on improving the fuel economy while guaranteeing that both the engine and the electric machine work in the high-efficiency area; typically, by taking into account knowledge of both bus routes and passenger loadings in a predictive manner. Selected examples of work in this direction can be found in \cite{Tianheng2015, Li2015, Li2016}. 

Our objective in this paper is to extend this line of inquiry further. Our basic setting is to consider a bus operator that has access to a fixed amount of renewable energy on a daily basis. For example, some operators may own solar farms or have access to wind generation. It makes sense to use this \underline{\em free  energy} before consuming electrical energy that is bought from the grid, and in situations where there is an oversubscription for this {\em free} energy, the operator then has a choice as to how this energy is distributed to each bus. For example, some drivers may be more efficient than others. Thus, it makes eminent economic sense, to distribute this free energy to reduce the impact of less efficient drivers in optimising the hybrid engine cycle, while at the same time ensuring that sufficient energy is consumed to make room for every unit of {\em free energy} that arrives the next time the buses recharge. Specifically, SPONGE for buses operates as follows. 
\begin{itemize}
	\item[A.] We introduce the forecast of generation of energy from renewable resources on a day ahead basis as a further variable to influence the energy management system for the bus operator.
	\item[B.] We use this forecast to prioritise the manner in which individual buses dissipate electrical energy.
	\item[C.] We do this by prioritising the utilisation of energy from renewable sources over other resources, and by taking account of the fact that some drivers/routes are more energy efficient than others.
\end{itemize}
Prioritising energy from renewable sources in this manner introduces a number of benefits for the bus operator and society.
\begin{itemize}
	\item The use of energy from renewable sources (e.g., wind turbines, dynamic water power, or solar power) achieves environmental health benefits with respect to the use of the ``power grid average'' electricity \cite{PNAS}.
	\item Financial benefits for the bus operator.
	\item Depleting PHEBs' batteries of a pre-specified quantity of energy allows better grid-demand balancing. That is, the energy provider knows in advance how much energy will be required by PHEBs, when connected for charging. This makes the electrical load of PHEBs to be fully predictable and dispatchable, thus mitigating the burden of the power grid to accommodate a not-known-in-advance electrical load.
\end{itemize}

Note that the proposed charging problem closely resembles the widely discussed practice of demand side management, where electricity customers shift their electrical loads taking into account the expected availability of energy from renewable sources (e.g., solar panels on the roofs of their houses). In fact, in this paper we are considering the possibility that buses accommodate the consumption of their batteries considering the amount of energy that will be available from renewable sources when recharging.

This paper extends previous work of some of the authors in \cite{Sponge_IJC} for the case of electric cars. While the basic idea of matching energy from renewable sources with space in the battery of the EVs remains the same, the case of PHEBs substantially differs from the case of Plug-in Hybrid Electric Vehicles (PHEVs) in several ways: (i) the a-priori knowledge of bus routes can be included in the formulation of the optimisation problem; (ii) we show that maximisation of $CO_2$ savings can be formulated as an optimisation problem as a function of the fraction of time that a PHEB spends in EV mode; (iii) the optimisation procedure can now be computed in a batch fashion, thus shifting the interest of optimisation algorithms from the time of convergence to other aspects, such as privacy preservation and communication costs of the algorithms. With this latter aspect in mind, our final contribution is to give a brief comparison of two competitive optimisation algorithms.

\section{Sponge Problem Formulation}
\label{Formulation}

\begin{figure}[htbp]
	\begin{center}
		{\includegraphics[width=3in, height = 2.3in]{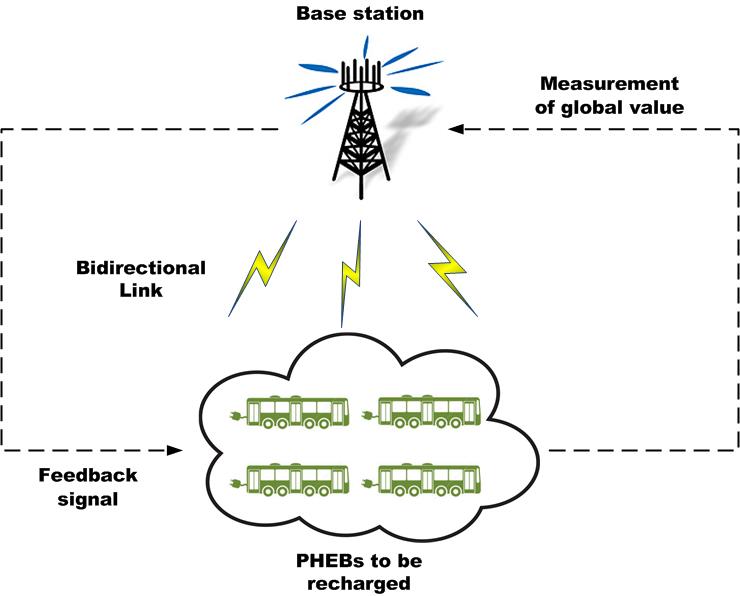}}
		\caption{A schematic diagram of the SPONGE programme}
		\label{feedbackLoop}
	\end{center}
\end{figure}

Let $\mathcal{N}=\left\{1,2,...,N\right\}$ denote the set of $N$ PHEBs participating to the SPONGE programme. We shall make the following assumptions:
\begin{itemize}
	\item
	We assume that after a number of trips along their (different) routes, the $N$ PHEBs stop for charging at the bus station. For instance, we can assume that the PHEBs will not drive from 11pm to 6am, and they will be charged in this time frame;
	\item
	We also assume that a 24-hour ahead forecast of energy from the renewable energy sources available to the operator will be available as well (e.g., a forecast of how much energy will be generated by the wind plants connected with the charging station at night time). We denote this amount of energy available by $E_{\textrm{av}}$;
	\item
	Early in the morning, before being dispatched along their routes, the buses will compute how the energy $E_{\textrm{av}}$ should be optimally shared among themselves during the day (i.e., in terms of energy consumption of their own batteries);
	\item
	In order to compute the optimal allocations of energy, we shall assume that each PHEB is equipped with a device to transmit messages to the central infrastructure via Vehicle-to-Infrastructure (V2I) technology;
	\item
	The central infrastructure has the ability to broadcast messages to the whole network of PHEBs using some Infrastructure-to-Vehicles (I2V) technology.
\end{itemize}
Note that in our set-up we shall not require vehicles to exchange information among themselves, and thus, we shall not require PHEBs to be equipped with Vehicle-to-Vehicle (V2V) communication devices. A schematic diagram of the above SPONGE paradigm is illustrated in Fig. \ref{feedbackLoop}.\newline

\begin{figure}[htbp]
	\begin{center}
		{\includegraphics[width=3in, height = 2.3in]{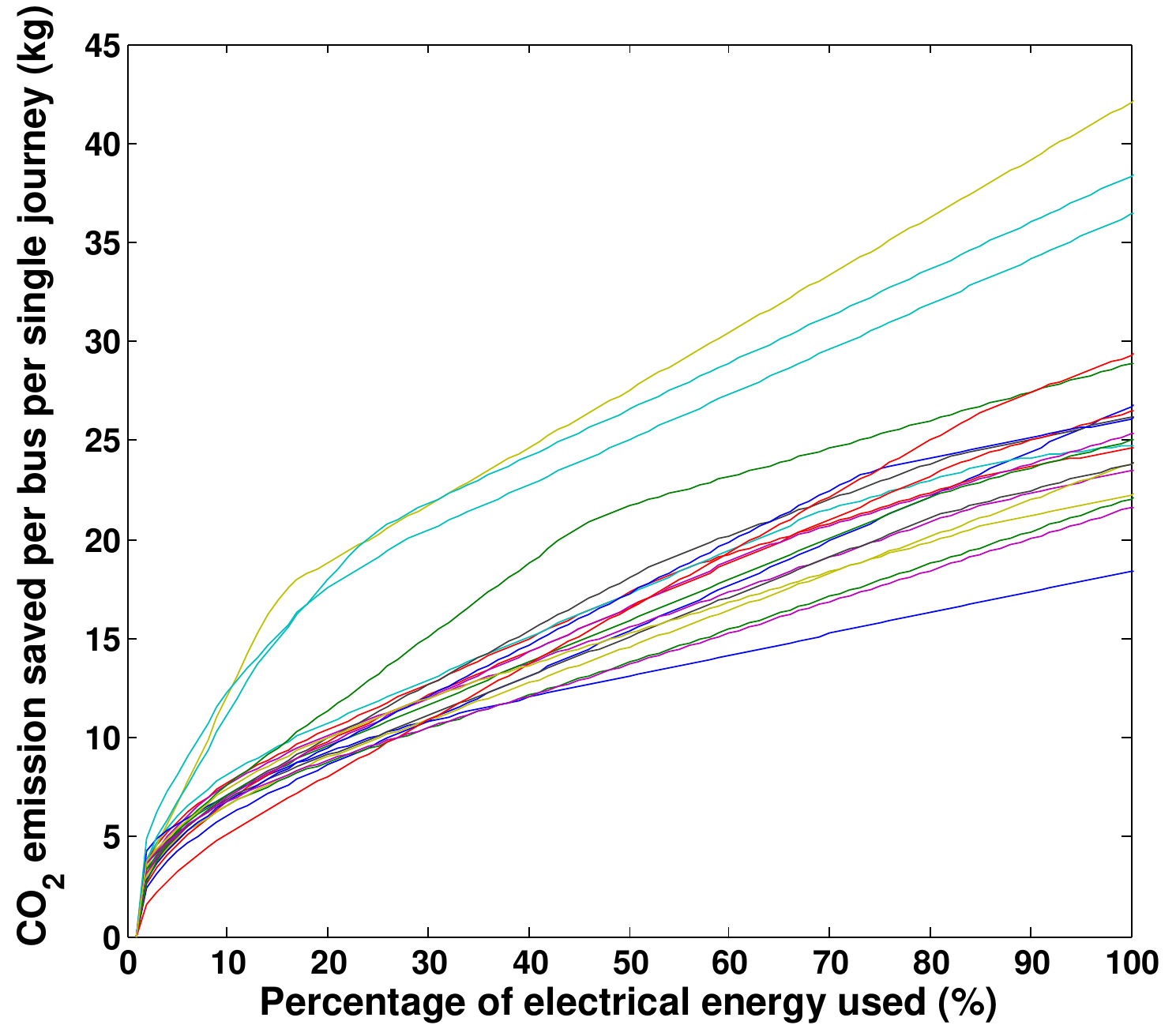}}
		\caption{Utility functions of 15 PHEBs in Dublin city. Note that some buses pollute more than others (and thus, have a greater potential in terms of $CO_2$ savings) depending on the characteristics of their routes (e.g., speed limits).}
		\label{UtilityFunctions}
	\end{center}
\end{figure}

When travelling along their routes, the buses will be able to choose when it is more convenient to switch from electric mode to ICE mode (i.e., using the Internal Combustion Engine) and back. In this context, we denote by $d_i$ the energy consumption by the $i'$th bus along its trip. Then we are interested in computing the solution of the following optimisation:
\begin{equation} 
\label{eq:opt}
\left\{
\begin {array}{l}
\underset{d_1,d_2,\ldots,d_{N}}{\max} \quad
\sum\limits_{i \in \mathcal{N}} f_{i}\left(d_i \right)\\
\\
{\text{s.t.}} ~
\sum\limits_{i\in \mathcal{N}} d_i = E_{\textrm{av}}\end{array}\right..
\end{equation}
In the optimisation problem (\ref{eq:opt}), the terms $d_i$ can be interpreted as a ``budget'' of energy that is allocated to the $i'$th bus in order to maximise a utility function of interest, such that the sum of the energy budgets allocated to all the buses matches $E_{\textrm{av}}$ as in the SPONGE spirit. Although in principle the utility function $\sum\limits_{i \in \mathcal{N}} f_{i}\left(d_i \right)$ may be chosen in an arbitrary fashion, to represent any {\em utility}, in this work we shall explore the particular case where one is interested in the utility of $CO_2$ emissions savings $f_{i}\left(d_i \right)$ as achieved by each bus. Clearly, each $f_{i}\left(d_i \right)$ is an increasing function of $d_i$ as no $CO_2$ emissions are saved when the bus travels all the time in ICE mode, while no pollution occurs when all the vehicles travel in electric mode all the time. The utility functions of 15 PHEBs that we shall study are shown in Fig. \ref{UtilityFunctions} as a function of the percentage of the use of the electrical engine for each bus. These functions are constructed from real data and the next section will explain how the utility functions are designed in detail. In addition, Section \ref{AIMD_Optimisation} will describe an attractive distributed solution of the optimisation problem (\ref{eq:opt}).

\subsection{Construction of the utility function}
\label{Utility_Functions}
\begin{figure}[htbp]
	\begin{center}
		\hspace{0.4cm}
		{\includegraphics[width=2.85in, height = 2.3in]{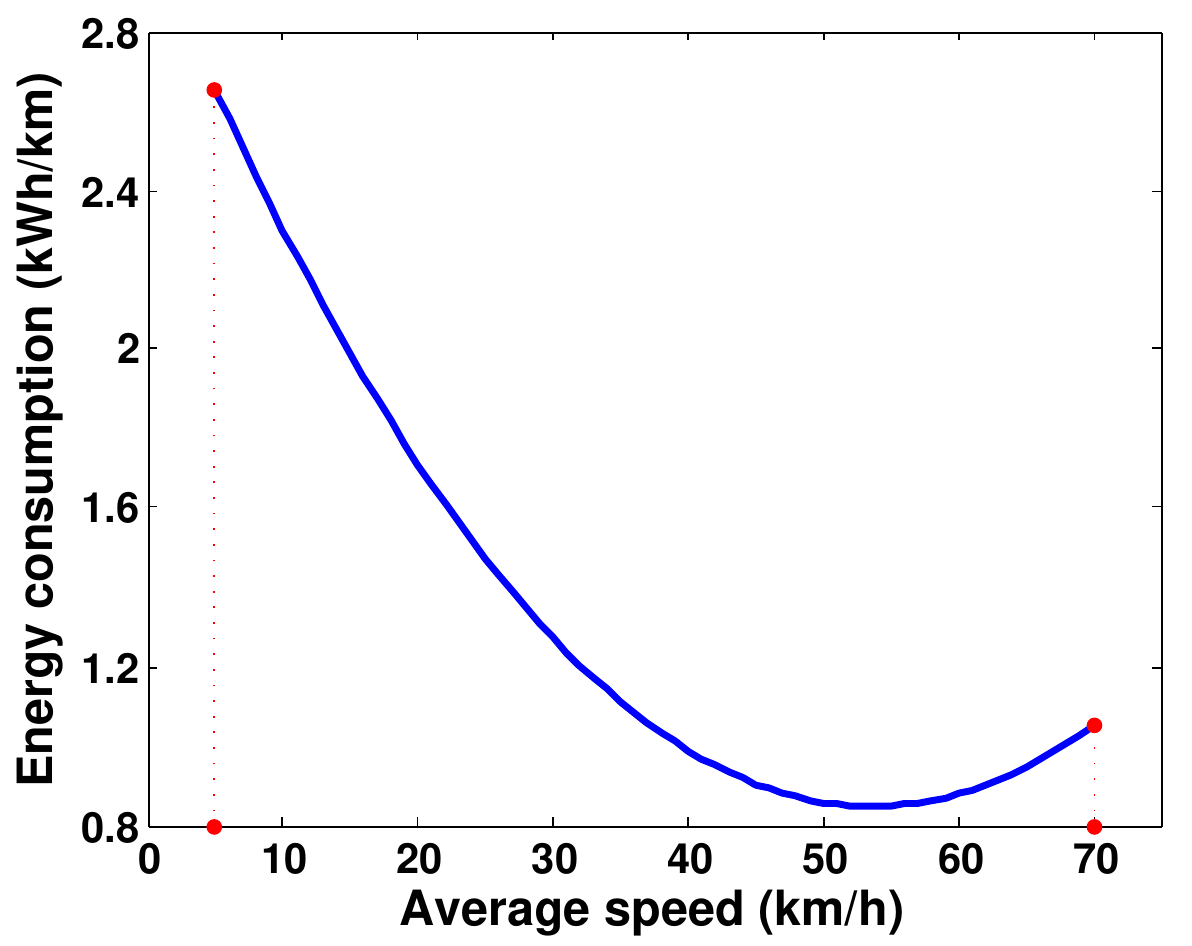}}
		\caption{A typical energy cost function for PHEBs.}
		\label{energyLosses}
	\end{center}
\end{figure}

\textbf{Electrical energy consumption}:
Under the assumption that a vehicle is travelling at a constant speed $s$, the amount of electrical energy consumption of vehicles can be modelled as a convex function of $s$, see \cite{vanHaaren,energyconsumption} for instance. The convex function depends in turn on, among other things, the physical characteristics of the bus. In our work, we used the real energy consumption data of a BYD electric bus \cite{energyconsumption} and noticed that it can be accurately approximated with a quadratic function of the vehicle speed $s$ as
\begin{equation}
e(s) = \alpha_0 s^2  + \alpha_1 s + \alpha_2.
\label{power_consumption}
\end{equation}
where $\alpha_0, \alpha_1, \alpha_2$ are all constant parameters. Using a conventional least square method to fit the real energy consumption we obtained $\alpha_0, \alpha_1 $ and $\alpha_2$ equal to $\frac{22}{28777}, \frac{-213}{2599}$ and $\frac{2384}{783}$, respectively, and the corresponding utility function is depicted in Fig. \ref{energyLosses}. In particular, note that the energy is large when the speed is large, which is caused by the fact that power consumption increases with the cube of the speed for aerodynamic reasons, and that the energy is large again when the speed is very low, due to the fact that travel times increase.

\begin{figure}[htbp]
	\begin{center}
		{\includegraphics[width=3in, height = 2.3in]{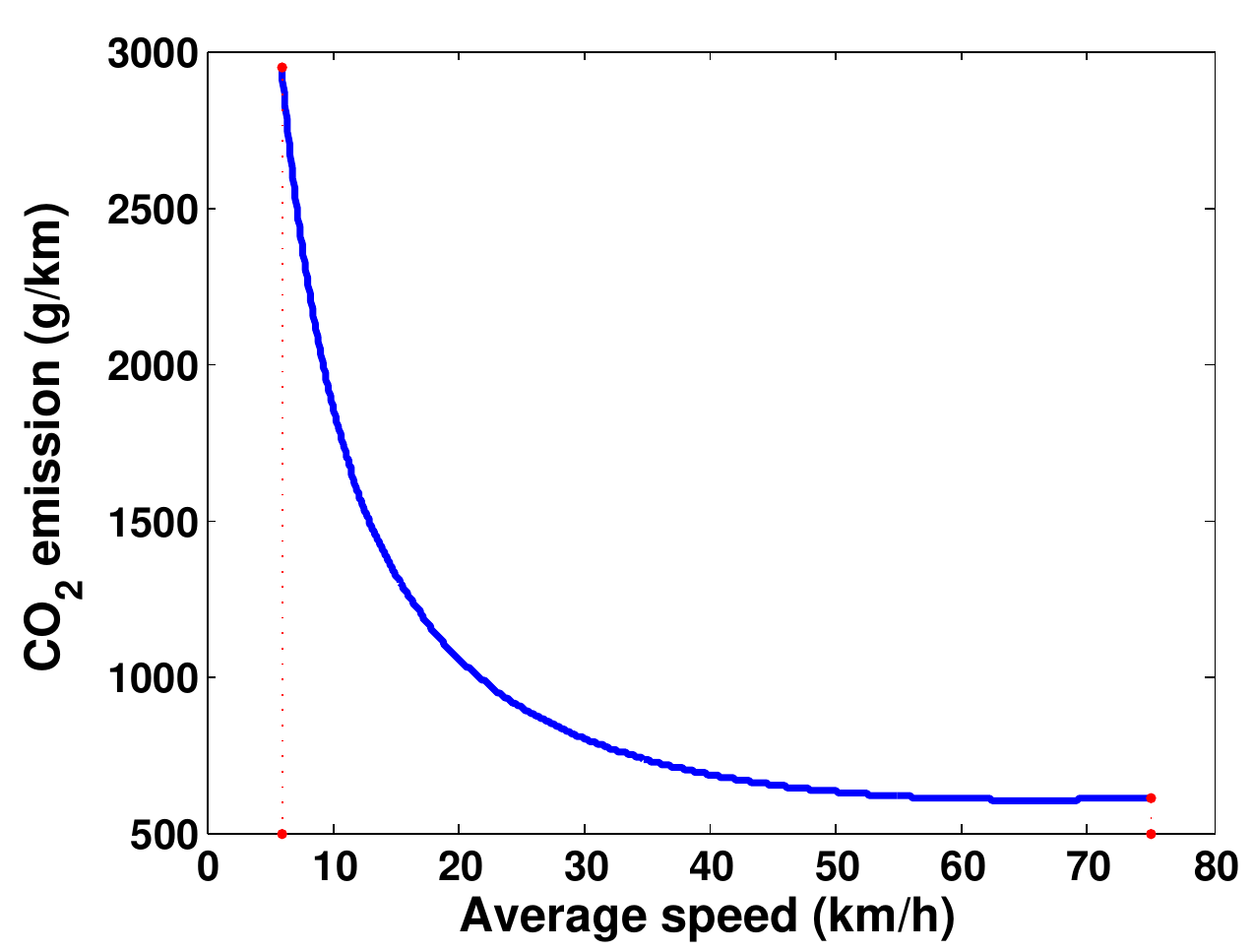}}
		\caption{$CO_2$ emission cost function for PHEBs.}
		\label{CO2emission}
	\end{center}
\end{figure}

\textbf{Saving of $\pmb {CO_2}$}: In an analogous manner to power consumption, $CO_2$ emissions may also be computed as a function of the speed of the vehicles, by adopting for instance the following well-known average-speed model (from \cite{emfactor}) 
\begin{equation}
\begin{gathered}
h(s)=
\text{k}\left(\frac{\text{a} + \text{b}s_{i}
	+ \text{c}s^{2}
	+ \text{d}s^{3}
	+ \text{e}s^{4}
	+ \text{f}s^{5}
	+ \text{g}s^{6}
}
{s
}\right),
\end{gathered}
\end{equation}
where $\text{a},\text{b},\text{c},\text{d},\text{e},\text{f},\text{g},\text{k} \in \mathbb{R}$ are used to specify different levels of emissions by different classes of vehicles. In particular, in our work we used the cost function depicted in Fig. \ref{CO2emission}, that corresponds to the vehicle code R203 in \cite{emfactor} (i.e., diesel buses with up to 15 tonnes of gross vehicle mass).
\newline

\textbf{Utility functions $\pmb{f_i}$}: As anticipated in Section \ref{Formulation}, the overall utility function $f_i(d_i)$ quantifies how much $CO_2$ has been saved by the $i'$th bus, provided that the bus is allowed to spend a budget of $d_i$ units of energy when traveling along its route. In the following, we shall assume that the whole path traveled by a bus during the day can be split into a number of very small sections, corresponding to the distance traveled by a bus in one second. For simplicity, we shall further assume that speed limits do not change within sections, and that a bus will travel at a speed equal to the speed limit within the section. Finally, we shall denote by $\mathcal{R}_i$ the set of all the sections traveled by the $i'$th PHEB, and by $\gamma_l$ the fraction of the time that a PHEB will travel in EV mode along the $l'$th section of the route. Due to the fact that all bus routes are fixed and known a priori, then the utility functions can be computed off-line by optimally deploying the energy budget $d_i$ along the routes as follows:
\begin{equation} \label{eq:opt1}
\left\{\begin{array}{l}
f_i(d_i) = \underset{\gamma_l}{\max} \quad \sum\limits_{l \in \mathcal{R}_i} h\left(s_{\textrm{max}}(l)\right) \cdot \gamma_l\\
\\
{\text{s.t.}} ~
\sum\limits_{l \in \mathcal{R}_i} e(s_{\textrm{max}}(l)) \cdot L(l) \cdot \gamma_l = d_i\\
\\
0 \leq \gamma_l \leq 1, l \in \mathcal{R}_i\end{array}\right.
\end{equation} 
where $L(l)$ denotes the length of the $l'$th section of a trip.
\begin{theorem}
	\label{Lemma1}
	The utility functions $f_i(d_i)$ are concave.
\end{theorem}

\noindent {\em Proof :} By definition \eqref{eq:opt1}

\begin{equation}
\left\{\begin{array}{l}
f_i(d_i) = \underset{\gamma_l}{\max} \quad \sum\limits_{l \in \mathcal{R}_i} h\left(s_{\textrm{max}}(l)\right) \cdot \gamma_l\\
\\
{\text{s.t.}} ~
\sum\limits_{l \in \mathcal{R}_i} e(s_{\textrm{max}}(l)) \cdot L(l) \cdot \gamma_l = d_i\\
\\
0 \leq \gamma_l \leq 1, l \in \mathcal{R}_i.\end{array}\right. . 
\end{equation}
Consider two variables $d_i^1 \neq d_i^2$, we can show that 
\begin{equation}
\left\{\begin{array}{l}
f_i(d_i^1) = \underset{\gamma_l}{\max} \quad \sum\limits_{l \in \mathcal{R}_i} h\left(s_{\textrm{max}}(l)\right) \cdot \gamma_l\\
\\
{\text{s.t.}} ~
\sum\limits_{l \in \mathcal{R}_i} e(s_{\textrm{max}}(l)) \cdot L(l) \cdot \gamma_l = d_i^1\\
\\
0 \leq \gamma_l \leq 1, l \in \mathcal{R}_i. \end{array}\right. ,
\end{equation}
Given a $0< \lambda < 1$, it then follows that
\begin{equation}
\left\{\begin{array}{l}
\lambda f_i(d_i^1) = \underset{\gamma_l}{\max} \quad \sum\limits_{l \in \mathcal{R}_i} h\left(s_{\textrm{max}}(l)\right) \cdot \lambda \gamma_l\\
\\
{\text{s.t.}} ~
\sum\limits_{l \in \mathcal{R}_i} e(s_{\textrm{max}}(l)) \cdot L(l) \cdot \gamma_l = d_i^1\\
\\
0 \leq \gamma_l \leq 1, l \in \mathcal{R}_i\end{array}\right. ,
\end{equation}
and 
\begin{equation}
\left\{\begin{array}{l}
\lambda f_i(d_i^1) = \underset{\gamma_l}{\max} \quad \sum\limits_{l \in \mathcal{R}_i} h\left(s_{\textrm{max}}(l)\right) \cdot  \bar{\gamma}_l\\
\\
{\text{s.t.}} ~
\sum\limits_{l \in \mathcal{R}_i} e(s_{\textrm{max}}(l)) \cdot L(l) \cdot \frac{\bar{\gamma}_l}{\lambda} = d_i^1\\
\\
0 \leq \frac{\bar{\gamma}_l}{\lambda} \leq 1, l \in \mathcal{R}_i.\end{array}\right. .
\end{equation}
Thus,
\begin{equation}
\label{eq5}
\left\{\begin{array}{l}
\lambda f_i(d_i^1) = \underset{\gamma_l}{\max} \quad \sum\limits_{l \in \mathcal{R}_i} h\left(s_{\textrm{max}}(l)\right) \cdot  \bar{\gamma}_l\\
\\
{\text{s.t.}} ~
\sum\limits_{l \in \mathcal{R}_i} e(s_{\textrm{max}}(l)) \cdot L(l) \cdot \bar{\gamma}_l = \lambda d_i^1\\
\\
0 \leq \bar{\gamma}_l \leq \lambda, l \in \mathcal{R}_i.\end{array}\right. .
\end{equation}
Similarly, we can show that  
\begin{equation}
\label{eq6}
\left\{\begin{array}{l}
(1-\lambda) f_i(d_i^2) = \underset{\gamma_l}{\max} \quad \sum\limits_{l \in \mathcal{R}_i} h\left(s_{\textrm{max}}(l)\right) \cdot  \hat{\gamma}_l\\
\\
{\text{s.t.}} ~
\sum\limits_{l \in \mathcal{R}_i} e(s_{\textrm{max}}(l)) \cdot L(l) \cdot \hat{\gamma}_l = (1-\lambda) d_i^2\\
\\
0 \leq \hat{\gamma}_l \leq (1-\lambda), l \in \mathcal{R}_i\end{array}\right. ,
\end{equation}
where adding \eqref{eq5} and \eqref{eq6} is equivalent to
\begin{equation}
\left\{\begin{array}{l}
\underset{\gamma_l}{\max} \quad \sum\limits_{l \in \mathcal{R}_i} h\left(s_{\textrm{max}}(l)\right) \cdot  (\bar{\gamma}_l+\hat{\gamma}_l)\\
\\
{\text{s.t.}} ~
\sum\limits_{l \in \mathcal{R}_i} e(s_{\textrm{max}}(l)) \cdot L(l) \cdot \bar{\gamma}_l = \lambda d_i^1\\
\\
\sum\limits_{l \in \mathcal{R}_i} e(s_{\textrm{max}}(l)) \cdot L(l) \cdot \hat{\gamma}_l = (1-\lambda) d_i^2\\
\\
0 \leq \bar{\gamma}_l \leq \lambda, l \in \mathcal{R}_i\\
\\
0 \leq \hat{\gamma}_l \leq (1-\lambda), l \in \mathcal{R}_i\end{array}\right. ,
\end{equation}
which is less or equal to 
\begin{equation}
\left\{\begin{array}{l}
\underset{\gamma_l}{\max} \quad \sum\limits_{l \in \mathcal{R}_i} h\left(s_{\textrm{max}}(l)\right) \cdot  (\bar{\gamma}_l+\hat{\gamma}_l)\\
\\
{\text{s.t.}} ~
\sum\limits_{l \in \mathcal{R}_i} e(s_{\textrm{max}}(l)) \cdot L(l) \cdot (\bar{\gamma}_l + \hat{\gamma}_l) = \lambda d_i^1 + (1-\lambda) d_i^2\\
\\
0 \leq \bar{\gamma}_l + \hat{\gamma}_l  \leq \lambda + (1-\lambda), l \in \mathcal{R}_i.\end{array}\right. .
\end{equation}
This implies that

\begin{equation}
\left\{\begin{array}{l}
\underset{\gamma_l}{\max} \quad \sum\limits_{l \in \mathcal{R}_i} h\left(s_{\textrm{max}}(l)\right) \cdot \gamma_l\\
\\
{\text{s.t.}} ~
\sum\limits_{l \in \mathcal{R}_i} e(s_{\textrm{max}}(l)) \cdot L(l) \cdot \gamma_l = \lambda d_i^1 + (1-\lambda) d_i^2\\
\\
0 \leq \gamma_l  \leq 1, l \in \mathcal{R}_i\end{array}\right.
\end{equation}
$= f_i(\lambda d^1_i+(1-\lambda)d^2_i)$. Therefore $\lambda f_i(d^1_i) + (1-\lambda) f_i(d^2_i) \leq f_i(\lambda d^1_i+(1-\lambda)d^2_i)$ and by definition the function $f_i$ is concave. ${\bf Q.E.D}$.\newline

\noindent \textbf{Remark:} Note that $\mathcal{R}_i$ is the set of all sections traveled by a bus during a day (or in general, between two different runs of the optimisation algorithm). Given that bus routes are typically cyclic, this implies that the same stretch of a road might appear more times in $\mathcal{R}_i$, and possibly with different values of optimal $\gamma_l$.\newline

\noindent \textbf{Remark:} Roughly speaking, the utility functions describe the maximum $CO_2$ savings that can be achieved by a bus, given that the bus is allocated a budget of $d_i$ units of energy that can be spent in travelling in electric mode. The knowledge of the route is used to optimally decide when it is best to drive in ICE mode and when in electric mode in order to maximise $CO_2$ savings without exceeding the energy budget. Built on such utility functions, the next section will show how to optimally allocate the expected energy $E_{av}$ into the single budgets.

\section{Algorithms and Optimal Solution}
\label{AIMD_Optimisation}

In principle, many different methods may be used to solve the optimisation problem \eqref{eq:opt} that arises in our work  (for instance ADMM-like algorithms \cite{boyd2011distributed}). ADMM (Alternating Direction Method of Multipliers) is a popular optimisation algorithm, that has been recently proposed as an evolution of other well-known optimisation algorithms, like the dual ascent and the method of multipliers. 
As an alternative to ADMM-like algorithms, our choice here is to adopt an AIMD-like algorithm \cite{wirth2014nonhomogeneous} to solve the problem in a distributed fashion. Such a choice is motivated by many reasons: 
\begin{itemize}
	\item
	\textbf{Low-communication requirements:} Although we have presented here a simple case study with a small number of buses, the same programme can be easily generalised to include hundreds of buses. Also, the batch optimisation formulation might be solved in real-time to account for non fully-predictable aspects (for example to respond to traffic peaks or weather forecast updates). In this context, it is convenient to consider the communication cost of solving the optimisation algorithm. {\em AIMD based optimisation can be solved using only intermittent binary feedback and can thus, unlike many other distributed optimisation techniques, be solved without the need to broadcast of the Lagrange multipliers in a pseudo-continuous manner}. \newline
	\item
	\textbf{Privacy-preservation requirements:} In our application, the utility functions $f_i(d)$ potentially reveal sensitive private information. For example, when formulated in a slightly different manner, these functions  may reveal how good a particular driver is on a given route. This information is potentially very useful for an employer and could potentially be used in a nefarious manner. In addition, in unionised environments, revealing these functions to an employer could also be of concern and consequently impede the adaptation of ideas like SPONGE. Given this context, a natural question is whether the distributed optimisation can be solved without revealing private information. {\em As we shall see, AIMD has some very nice privacy properties}. \newline
	\item
	\textbf{Agent actuation:}  {\em AIMD requires very little actuation ability on the agent-side}. This is in contrast to ADMM where at each time step, agents must solve a local optimisation problem.\newline
	\item
	\textbf{Algorithm parameterisation:} {\em In AIMD the gain parameters of the network are independent of network dimension; rather, they only depend on the largest derivative over all utility functions}. Thus, selecting a gain for the algorithm is extremely simple in the case of AIMD.
\end{itemize}
As we shall further discuss in the following section, AIMD is thus a convenient alternative to ADMM, when the previous aspects are relevant.

\subsection{AIMD Algorithm}

Additive Increase Multiplicative Decrease (AIMD) algorithms were originally applied for solving issues arising in network congestion in the Internet \cite{rejaie1999rap}. To date, this idea has been widely explored for the design of practical algorithms for other applications as well, as for instance, network applications see \cite{corless2012ergodic, budzisz2009strategy, shorten2005analysis}, and smart grid applications see \cite{liu2013investigation, crisostomi2014plug, studli2014optimal}. More recently, an unsynchronised AIMD algorithm based on the nonhomogeneous place-dependent Markov chains model was proposed in \cite{wirth2014nonhomogeneous} to solve utility optimisation problems. The pseudo-code of the proposed algorithm is given in Algorithm~\ref{alg:Algorithm1}.
\begin{algorithm}[htbp]
	\caption{Unsynchronised AIMD Algorithm}
	\begin{algorithmic}[1]	
		\State \textbf{Initialisation:} $k = 1$, $d_i(k) = 0$; 
		\State Broadcast the parameter $\Gamma$ to the entire networks;
		\While{$k < k_{\textrm{max}}$}
		\If {$\sum_{i=1}^{N} d_i(k) < E_{\textrm{av}}$}
		\State $d_i(k+1) = d_i(k) + \alpha$
		\ElsIf {with probability $p_i(k) = \Gamma \frac{1}{ \overline{d_i}(k) f_i'(\overline{d_i}(k))}$}
		\State $d_i(k+1) = \beta d_i(k)$
		\Else 
		\State $d_i(k+1) = d_i(k) + \alpha$  
		\EndIf 	
		\State $k = k +1$
		\EndWhile
	\end{algorithmic}
	\label{alg:Algorithm1}
\end{algorithm}
Note that the algorithm does not compute the optimal budgets $d_i$ in a single step, but in an iterative fashion, as $d_i(k)$ represents the value of the unknown energy to be allocated to the $i'$th PHEB, computed at time step $k$. For large values of $k$, $d_i(k)$ will eventually converge to the optimal solution that maximises the environmental benefits (while still satisfying the energy constraint). In Algorithm 1, $k_{\textrm{max}}$ represents the maximum number of iterations before the algorithm stops (e.g., after five minutes of iterations).

The basic idea of Algorithm \ref{alg:Algorithm1} is that if the sum of the $d_i(k)$ of all PHEBs is smaller than $E_{av}$, then each PHEV increases its target energy consumption $d_i(k)$ at the next iteration $k + 1$ by a quantity $\alpha$. However, if the sum of the energy budgets of all PHEVs exceeds $E_{av}$ (this situation is usually called as a congestion event), then each PHEB decreases its energy consumption by a multiplicative factor $0 < \beta < 1$ with probability $p_i(k) = \Gamma \frac{1}{\overline{d_i}(k) f_i'(\overline{d_i}(k))}$, where $\Gamma$ is a constant common broadcast parameter, and $\overline{d_i}(k)$ is the time average of the sequence of $d_i(k)$ at congestion events, up to the last iteration. It is proved in \cite{wirth2014nonhomogeneous} that $\overline{d_i}(k)$ approaches to the optimal solution of the problem when Algorithm 1 converges and where the optimisation is carried out over the $f_i(\bar{d}_i)$. \\

\noindent \textbf{Remark (AIMD):} The philosophy underlying the AIMD algorithm is to adjust $p_i(k)$ and $d_i(k)$ at every time step $k$ such that for large values of $k$, $f_i'(\overline{d_i}(k)) = f_j'(\overline{d_j}(k)), ~\forall i \neq j \in \mathcal{N}$, or in other words the PHEBs achieve consensus on the derivatives of their utility functions. This, with strict convexity of the utility functions, is both necessary and sufficient for optimality when feasibility is guaranteed. This property is known from elementary optimisation theory. Algorithm \ref{alg:Algorithm1} was originally designed in \cite{wirth2014nonhomogeneous} to \textit{minimise} a cost function of interest, here we slightly adapt it to \textit{maximise} $CO_2$ savings. Accordingly, given that each utility function $f_i$ in our case is strictly concave, and that the $p_i$ are strictly non-increasing in our problem, then we can adapt the algorithm in \cite{wirth2014nonhomogeneous} so that consensus is achieved on $1/f_i'$, and the convergence and optimality properties of the algorithm are preserved.\newline

\noindent \textbf{Remark (Privacy):} We now make some brief comments concerning the privacy properties of AIMD based optimisation. Recall that we assume that the central agent may receive the value $d_i$ from agent $i$, and performs the aggregation $A = \sum_{i=1}^{N} d_i$. We also assume that there are no incentives for an agent to cooperate with the central agent to help deduce the $f_i'$; that is, all agents, other than the central agent, are honest.  Given this basic setting, one may discern the following four basic levels of privacy.
\begin{itemize}
	\item[(i)] \underline{Absolute utility privacy (AUP) :} Here, the central agent cannot deduce $f_i(d)$ based on knowledge available to it. This is a basic level of privacy.
	\item[(ii)] \underline{Relative utility privacy (RUP) :} Here the central agent cannot deduce whether  $f_i(d) > f_j(d)$. This again, is a basic level of privacy. 
	\item[(iii)] \underline{Absolute derivative privacy (ADP) :} Here, the central agent cannot deduce $f'_i(d)$ based on knowledge available to it. This information is important since it allows the central agents to estimate the {\em price} elasticity of individual agents.
	\item[(iv)] \underline{Relative derivative privacy (RDP) :} Here the central agent cannot deduce whether  $f'_i(d) > f'_j(d)$.
\end{itemize}
Privacy preservation is beyond the scope of this paper. However, we note briefly that the stochastic AIMD algorithm allows us to give  guarantees regarding some of these privacy categories. First since the optimisation is based on $f'_i(d_i)$, the AIMD algorithm can be considered \emph{AUP}- and \emph{ADP}-private. Deducing any $f'_i(d_i)$ would require estimation of the $p_i(k)$ in Algorithm \ref{alg:Algorithm1}. Clearly, this is difficult (but not impossible) except at optimal points. However, since our algorithm only requires an implicit consensus among all derivatives, one may replace in the formula for $p_i(k)$ (i.e., line 6 of Algorithm \ref{alg:Algorithm1}), $f'_i(d_i)$ with $g(f'_i(d_i))$, where $g$ is chosen so that the convergence conditions in \cite{wirth2014nonhomogeneous} are satisfied. Clearly, without knowledge of the function $g$, the central agent cannot deduce $f'_i(d_i)$ even if the probabilities $p_i$'s are correctly estimated when the algorithm converges.

\section{SUMO Simulations}

\subsection{Simulation Set-up}

\begin{figure}[htbp]
	\begin{center}
		{\includegraphics[width=3in, height = 2.3in]{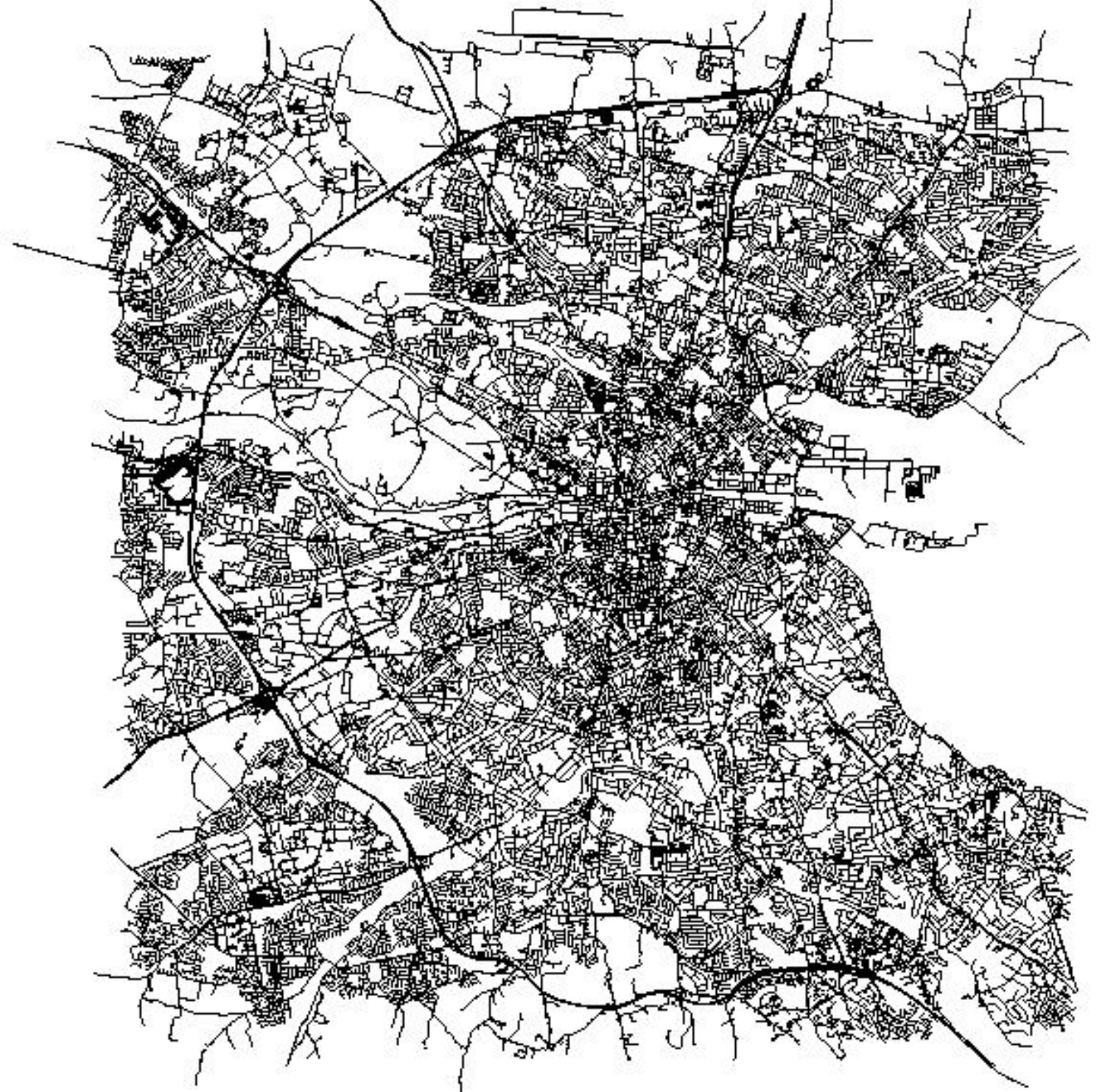}}
		\caption{Road network of Dublin City, Ireland, imported from OpenStreetMap, used in our simulations.}
		\label{DublinMap}
	\end{center}
\end{figure}

In this section, we evaluate the performance of the proposed AIMD algorithm in a realistic traffic scenario, where vehicular flows are simulated using the popular mobility simulator SUMO \cite{SUMORef}. In doing so, we shall also compare the results obtained using AIMD with those obtained with the ADMM algorithm. Finally, we shall also shortly discuss the impact of inaccurate predictions made by weather forecasting tools (e.g., from solar PV panels). All the simulations are performed over the road network of Dublin, Ireland, depicted in Fig. \ref{DublinMap}, imported from OpenStreetMap \cite{openstreetmap}. 

\subsection{Simulation Results}

\begin{figure}[htbp]
	\begin{center}
		\hspace{-0.1cm}
		{\includegraphics[width=3.1in, height = 2.3in]{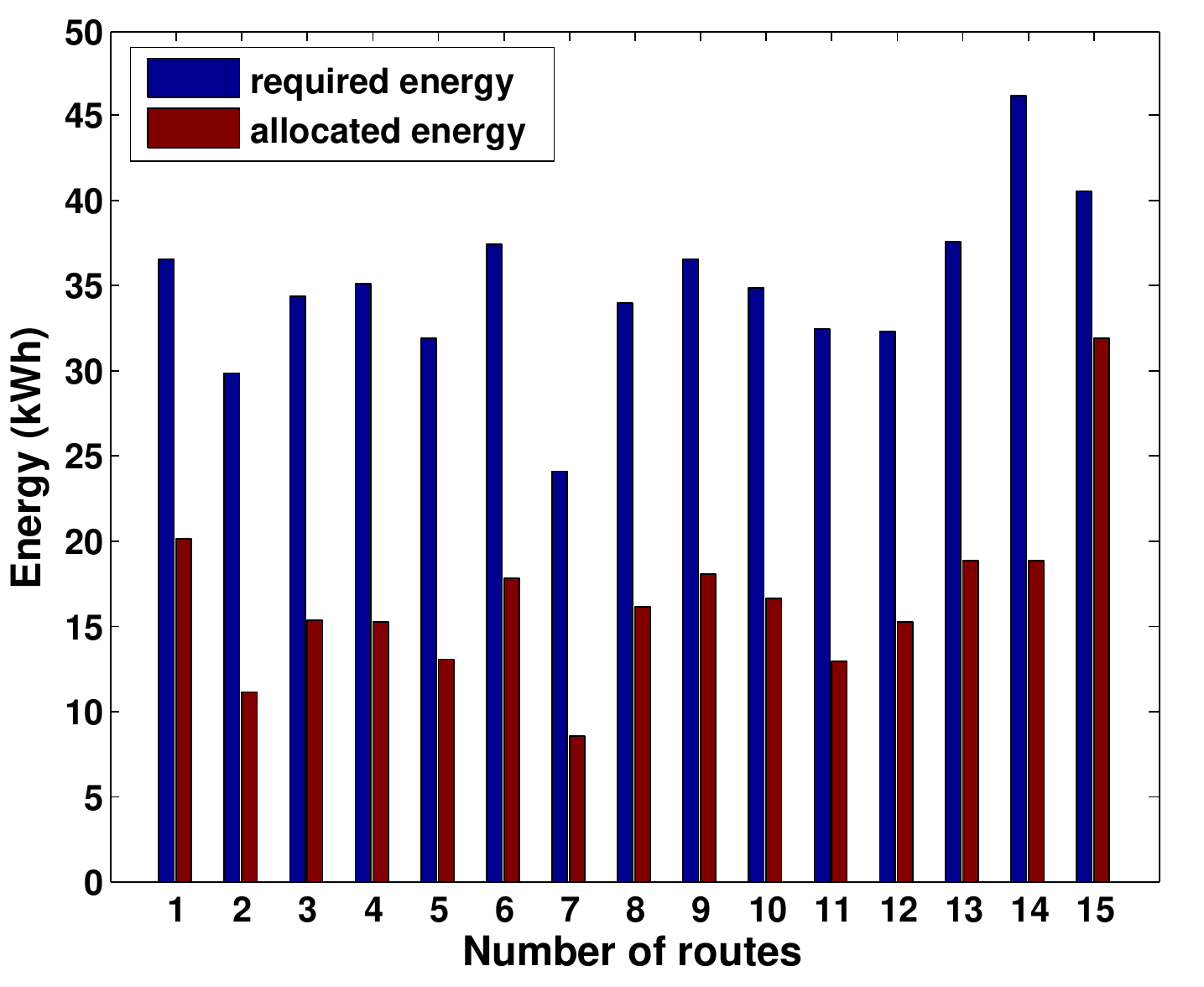}}
		\caption{Comparison of the required energy and the optimal allocated energy for 15 bus trips}
		\label{total_energy}
	\end{center}
\end{figure}

\begin{figure}[htbp]
	\begin{center}
		{\includegraphics[width=3.1in, height = 2.3in]{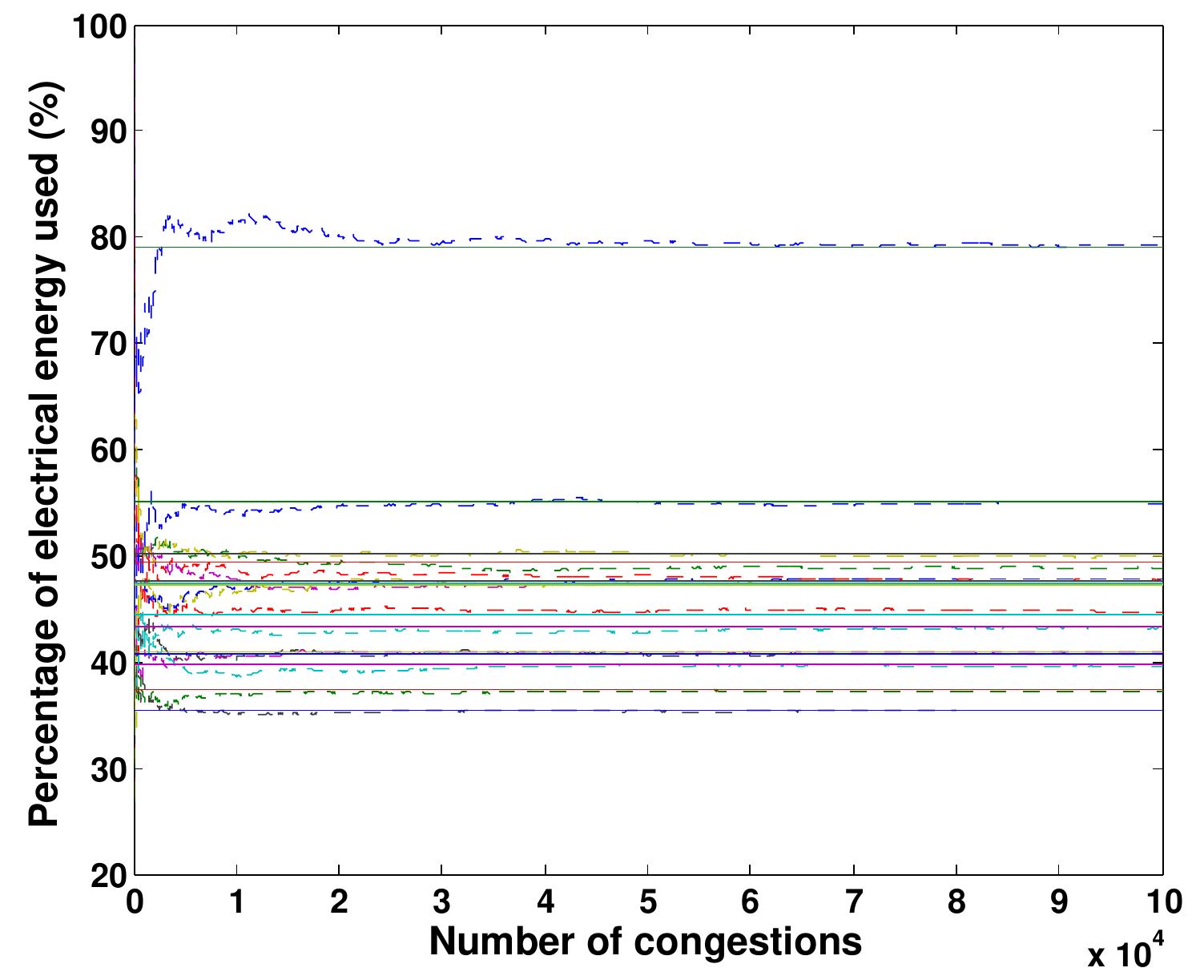}}
		\caption{Evolution of AIMD}
		\label{AIMD1}
	\end{center}
\end{figure}

\begin{figure}[htbp]
	\begin{center}
		\hspace{-1cm}
		{\includegraphics[width=3.2in, height = 2.3in]{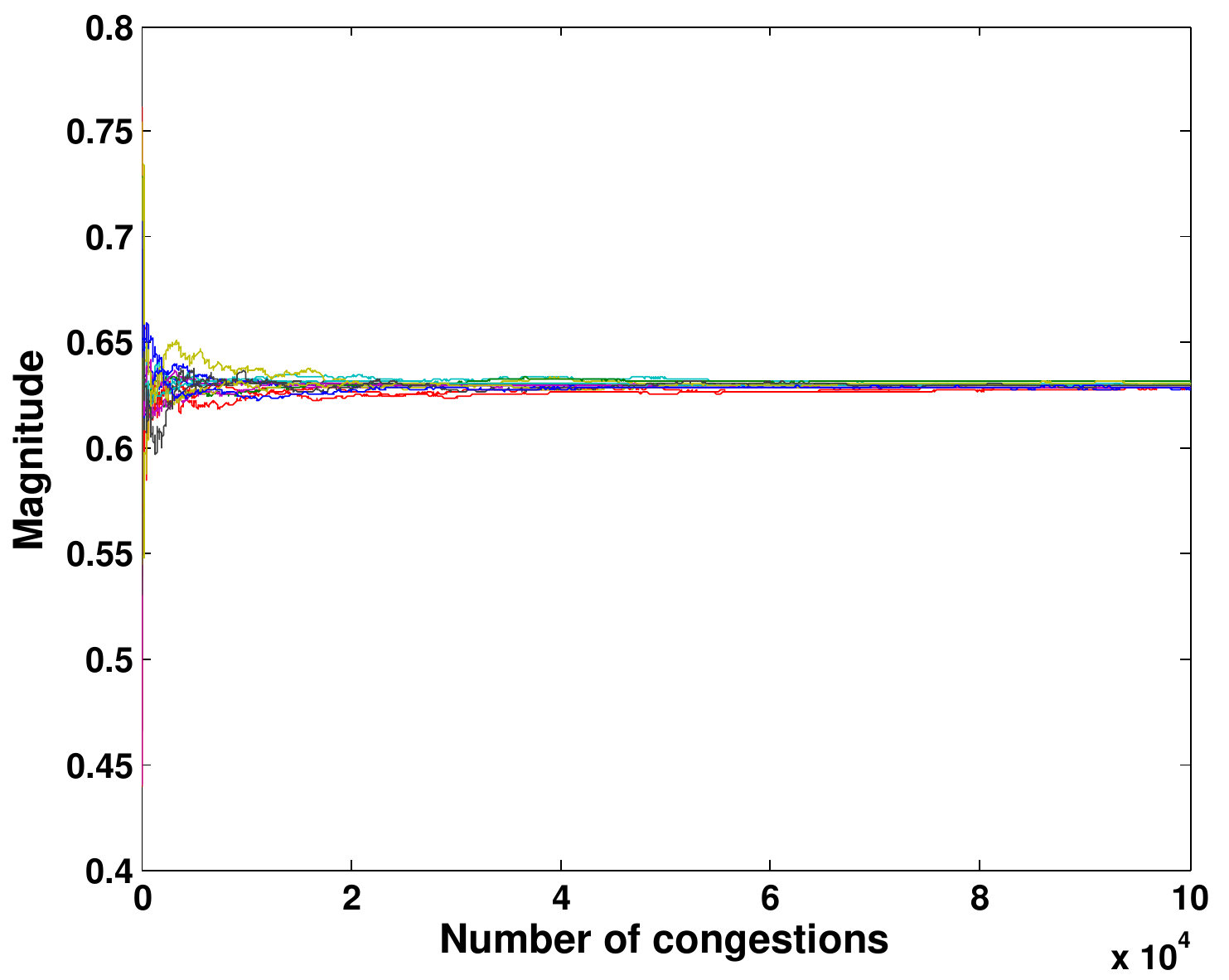}}
		\caption{Evolution of $f_i'(\overline{d_i}(k))$s converges to consensus}
		\label{AIMD2}
	\end{center}
\end{figure}

\begin{figure}[htbp]
	\begin{center}
		\hspace{-1cm}
		{\includegraphics[width=3.2in, height = 2.3in]{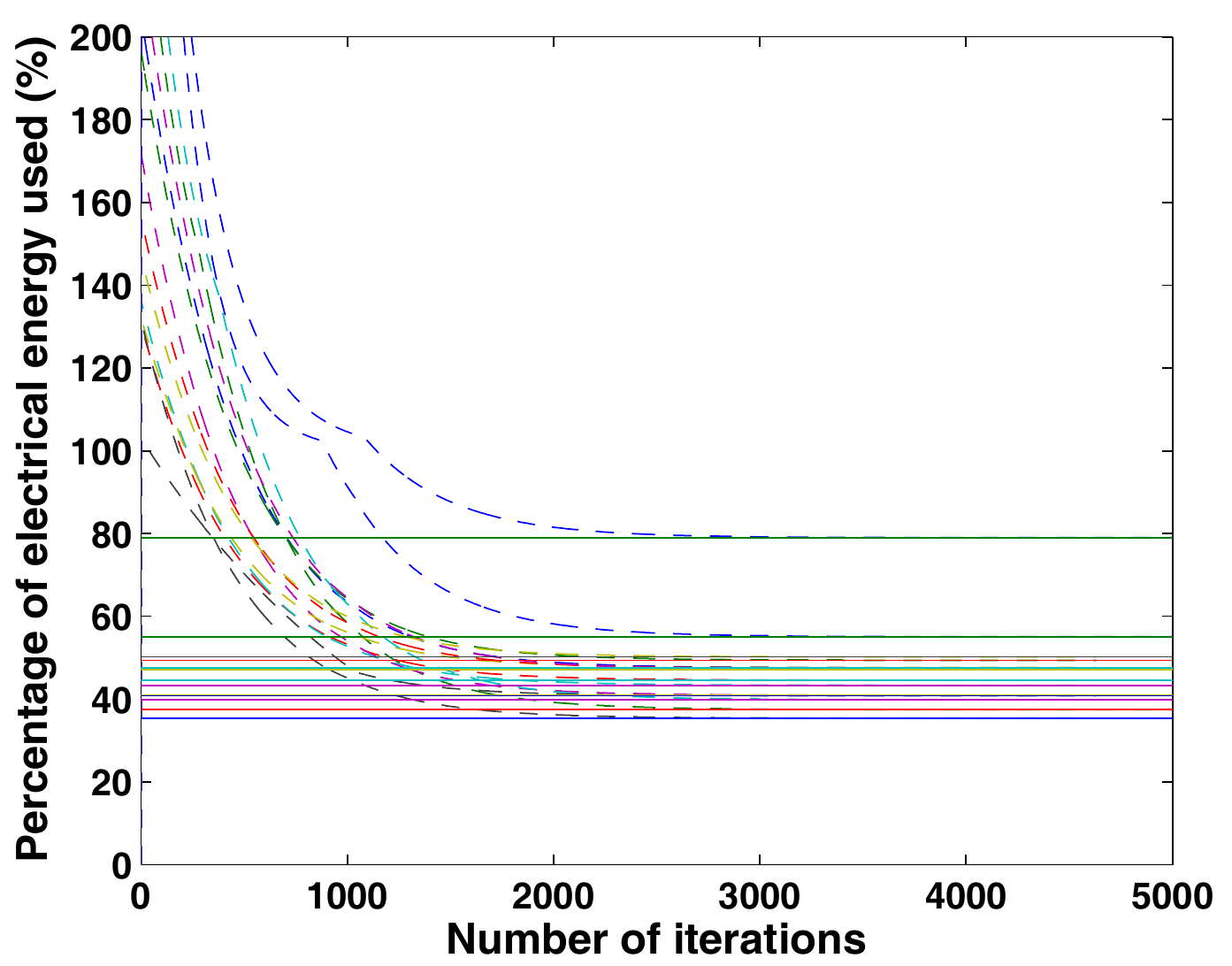}}
		\caption{Evolution of the distributed ADMM}
		\label{ADMMresults}
	\end{center}
\end{figure}

We assume that 15 PHEBs participate in a SPONGE programme in Dublin city, Ireland. We further assume that weather forecasting tools predict an availability of 250 kWh in the next charging period. Ten minutes before starting their routes, the buses and the CA solve the optimisation problem using the described AIMD algorithm, and optimally allocate the 250 kWh of available energy to 15 different bus trips. Fig. \ref{total_energy} compares the overall energy that would be required for each of the 15 trips when travelling the full route in EV mode (blue bars) with the optimal allocated budgets (red bars). Fig. \ref{AIMD1} shows that the AIMD algorithm converges to the optimal solution that can be easily computed by solving \eqref{eq:opt} using a standard nonlinear optimisation solver. Fig. \ref{AIMD2} reveals that the necessary condition for optimality (KKT) has been achieved when AIMD converges (i.e., the derivatives of the utility functions converge to the same value). Comparatively, Fig. \ref{ADMMresults} demonstrates that ADMM converges to the same solution as AIMD. We note that although ADMM requires less iterations to converge (around 3,500 iterations) compared to AIMD (around 20,000 iterations), ADMM requires more data to be transmitted to the agents. For instance, we assume that at each iteration ADMM needs to broadcast a packet (with multipliers) to all buses in 32 bits, therefore the total data required for algorithm convergence is $32 \times 3500 = 14$kB. On the other hand, AIMD needs to transmit one bit for all buses only on congestion events so the maximum data that is transmitted is less than $4$kB. This shows that AIMD is competitive from the perspective of communication overhead when compared to the ADMM algorithm.

\begin{figure}[htbp]
	\begin{center}
		{\includegraphics[width=2.9in,height = 2.3in]{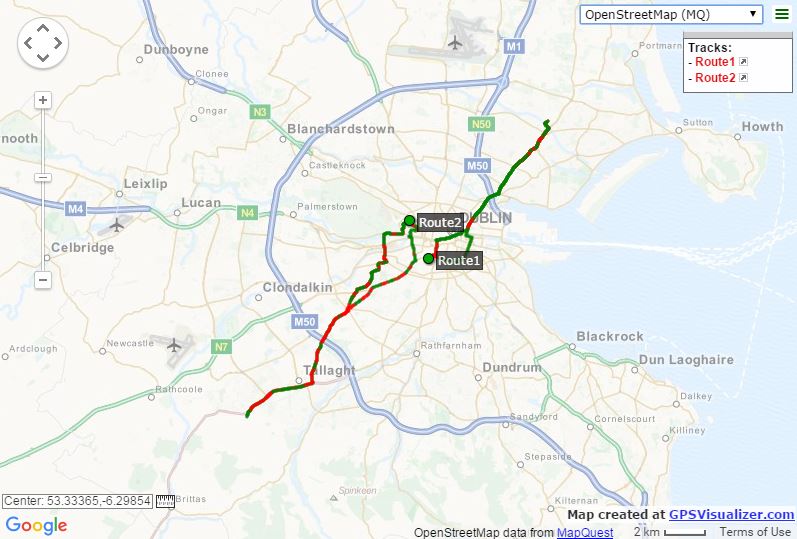}}
		\hspace{0.2cm}
		\caption{Electrical energy distribution of two bus routes when $E_{av} = 250$kWh. Green lines indicate PHEBs driving in EV mode and red lines indicate PHEBs in ICE mode}
		\label{Eav250}
	\end{center}
\end{figure}

\begin{figure}[htbp]
	\begin{center}
		\vspace{0.5cm}
		{\includegraphics[width=2.9in,height = 2.3in]{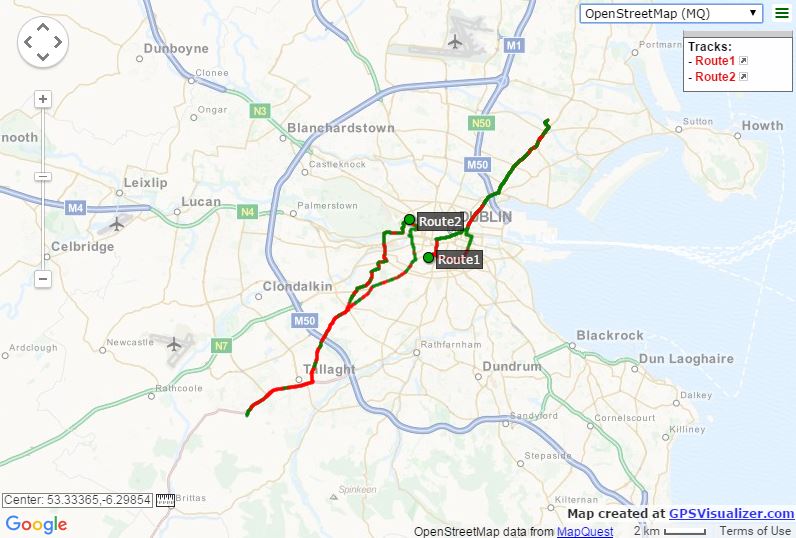}}
		\caption{Electrical energy distribution of two bus routes when $E_{av} = 100$kWh}
		\label{Eav100}
	\end{center}
\end{figure}

\begin{figure}[htbp]
	\begin{center}
		\vspace{0.25cm}
		{\includegraphics[width=2.9in,height = 2.3in]{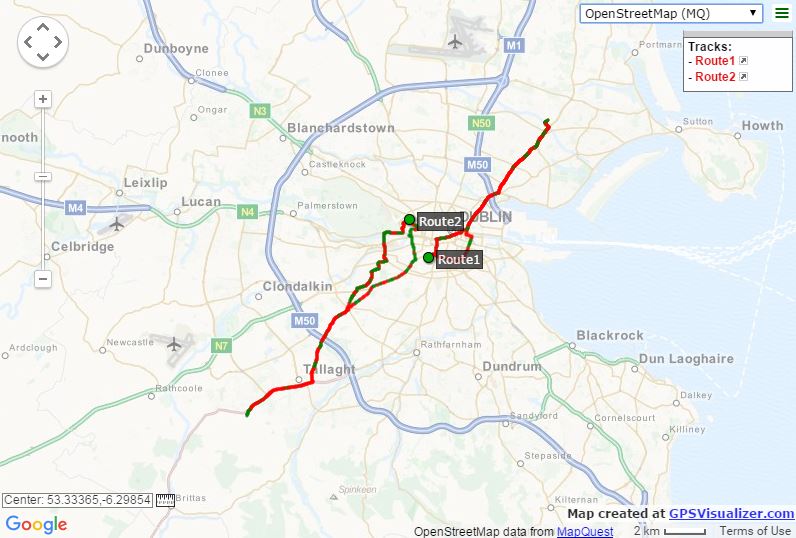}}
		\caption{Electrical energy distribution of two bus routes when $E_{av} = 50$kWh}
		\label{Eav50}
	\end{center}
\end{figure}

Figs. \ref{Eav250}-\ref{Eav50} illustrate the sections of the road where ICE and EV modes are used for buses 1 and 2 in a single journey for three different scenarios of total available energy $E_{av}$: 250kWh, 100kWh, and 50kWh. Clearly, in scenarios where the available energy is lower, the green sections corresponding to EV mode are shorter than the scenarios where the available energy is larger. When $E_{av} =250$kWh the total savings of $CO_2$ emissions for the 15 bus routes is 186.6kg, while the total savings is 83.37kg and 44.30kg for the cases where $E_{av} = 100 $kWh and $E_{av} = 50 $kWh, respectively. 

\section{Comments on the practical implementation of SPONGE}

To conclude the paper we now briefly comment on testing and implementation of the SPONGE algorithm.\newline

{\em A. Large-Scale Traffic Simulator} As we have mentioned, all simulations are based on the SUMO simulation environment. SUMO \cite{SUMORef} is an open source, microscopic road traffic simulation package primarily being developed at the Institute of Transportation Systems at the German Aerospace Centre (DLR). SUMO is designed to handle large road networks, and comes with a ``remote control'' interface, TraCI (short for Traffic Control Interface) \cite{b4}, that allows one to adapt the simulation and to control singular vehicles on the fly.\newline 

\begin{figure}[t]
	\centering
	\includegraphics[width=3.0in]{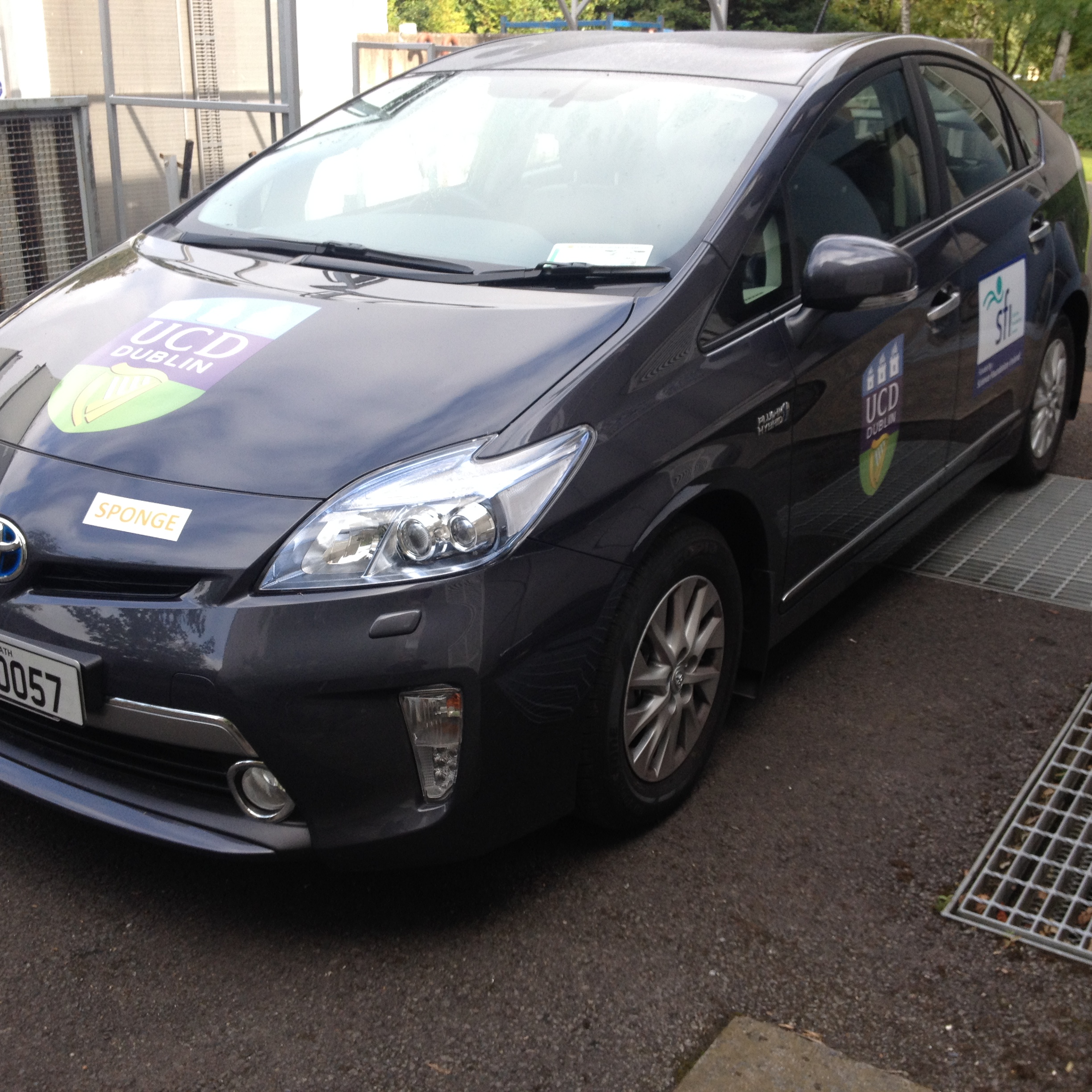}
	\caption{Field-test vehicle:  2015 Toyota Prius.}\label{f1}
\end{figure}

{\em B.  Test Vehicle :} While we have not yet implemented SPONGE in a real bus, the algorithm has been implemented  in a real test vehicle.  Our test vehicle is a 2015 Toyota Prius VVTi 1.8 5DR CVT Plugin Hybrid vehicle and is pictured in Fig. \ref{f1}.  The engine management system of the Prius allows the vehicle to be powered by the ICE alone, the battery, or using a combination of both, and it is this degree of freedom that we exploit to implement SPONGE.  For the purpose of this programme, we have made some important modifications to the basic vehicle to make it behave as a context-aware vehicle. First, we automate the switching of the vehicle from ICE to EV mode by adapting the ÒEVmode buttonÓ hardware in the vehicle. For this purpose, a dedicated Bluetooth-controlled mechanical interface was constructed to override the manual EV button based on signals from a smartphone. The switching is based on GPS location, external context information, and onboard signals such as speed and battery level. Second, special-purpose hardware was constructed to permit communication between a smartphone and the controller area network (CAN) bus. The Prius provides a CAN access on the vehicle diagnosis {\em On Board Diagnosis II (OBDII) interface}. Our hardware module acts as a gateway between this CAN interface and the smartphone.
The module is directly connected to CAN and to the smartphone via Bluetooth.  Communication to other vehicles, to GPS, and to a cloud server is also realized using a smartphone device. To control the driving mode, the software connects via Bluetooth to a mechanical switch to
toggle driving mode between the EV mode and non-EV driving modes.  In our application we use a Samsung Galaxy S III mini (model no. GT-I8190N) running the Android Jelly Bean operating system (version 4.1.2) and the OBD2 interface device that we used was the Kiwi Bluetooth OBD-II Adaptor by PLX Devices.\footnote{PLX Devices Inc., 440 Oakmead Parkway, Sunnyvale, CA 94085, USA.  Phone:  +1 (408) 7457591.  Website:  \url{http://www.plxdevices.com}}. \newline

\begin{figure}[htbp]
	\begin{center}
		{\includegraphics[width=3in, height = 2.1in]{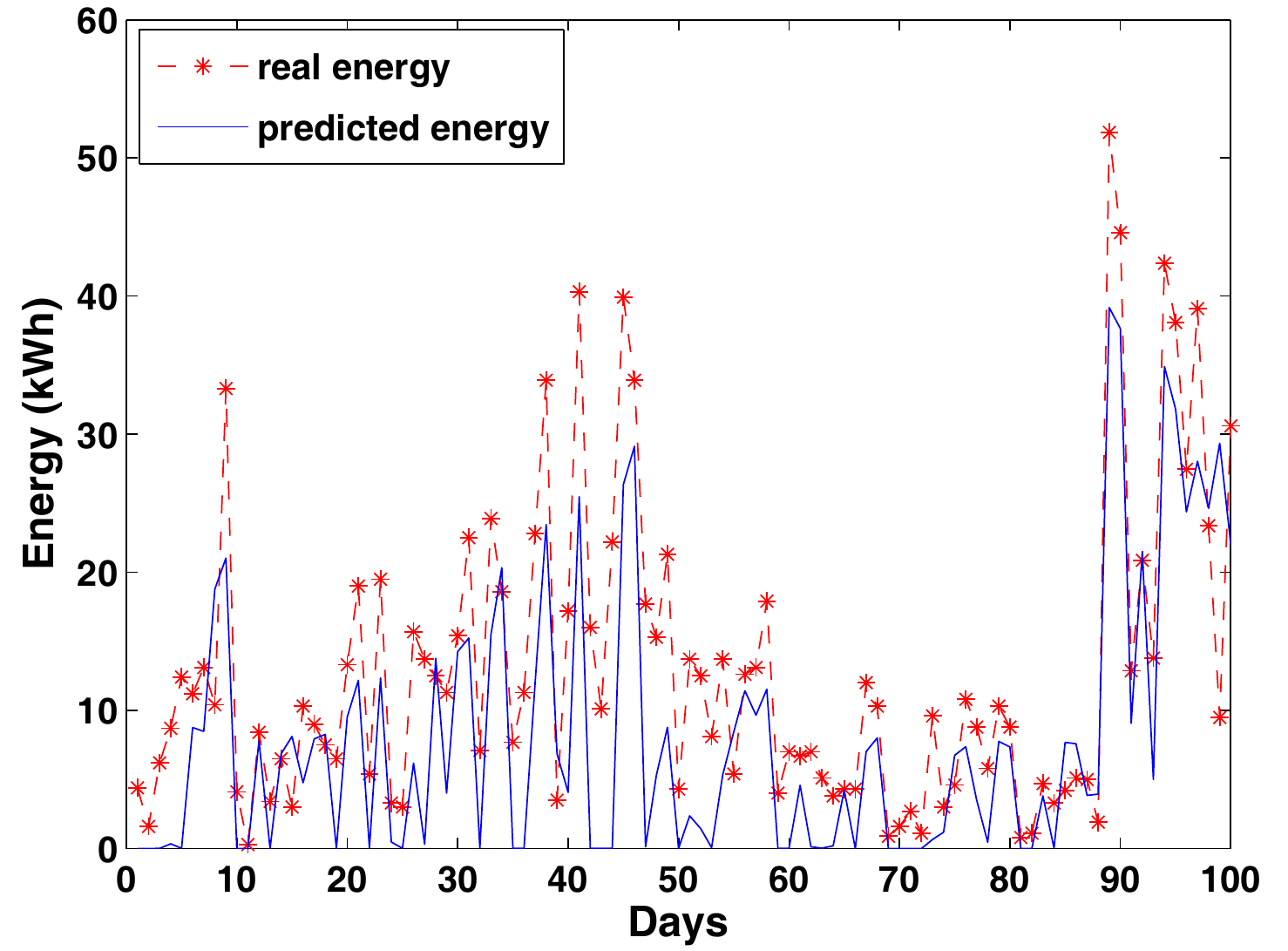}}
		\caption{Comparison between the real and the predicted energy generated from PV panels in UCD}
		\label{UCDNMAEbias}
	\end{center}
\end{figure}

\begin{figure}[htbp]
	\begin{center}
		{\includegraphics[width=3in, height = 2.1in]{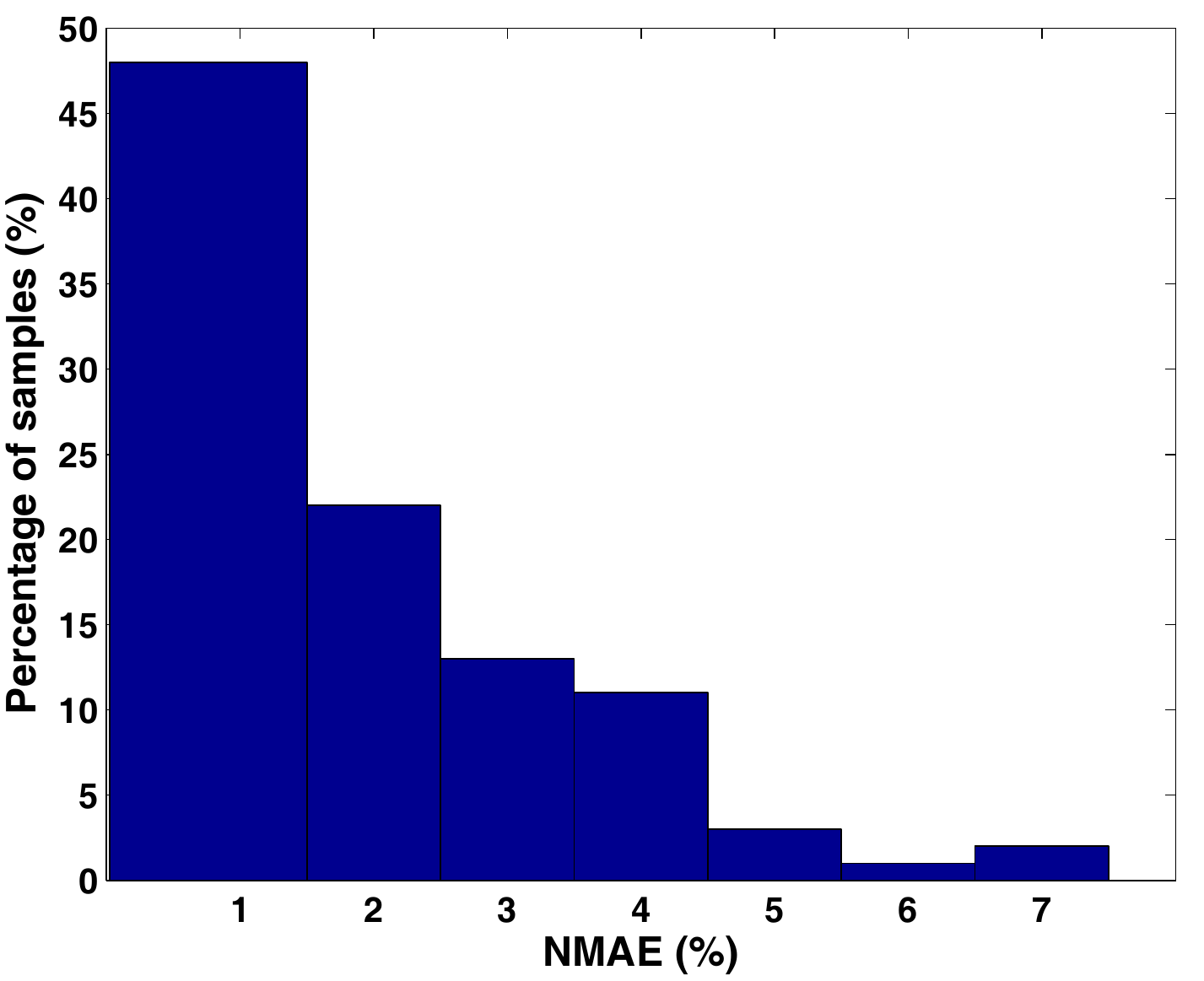}}
		\caption{Histogram of the percentage of NMAE}
		\label{UCDNMAE}
	\end{center}
\end{figure}

{\em C. Weather forecasting :} An important component in any real practical implementation of the SPONGE programme is the ability to have a reasonably accurate, and cheap, prediction of the expected energy that will be available for charging $E_{\textrm{av}}$. To obtain a feeling for fidelity of such tools, we evaluated the accuracy of a {\em free}  online forecasting tool over a 3 month period. The tool that we evaluated is provided by the Technical University of Crete and is described in \cite{energyPredication}, where the energy generated by a solar plant can be predicted (anywhere in the world) by simply providing the technical parameters of the plant. We collected real data on-site from PV panels mounted on the flat roof of the building in University College Dublin, Ireland. We recorded a total of 100 days and the predicted and the actual recorded energy are shown in Fig. \ref{UCDNMAEbias}. As also shown in Fig. \ref{UCDNMAE} the predictions are relatively accurate with 80\% of the predictions within 3\% of Normalised Mean Absolute Error (NMAE) and the maximum NMAE is 7\%. Thus, our data suggests that accurate predictions can be performed even for small powers, and even when a free online tool is employed. As for wind power forecasts, we note that a recent study in Germany reported that ``typical wind-forecast errors for representative wind power forecasts for a single wind project are $10\%\ -15\%$ root mean square error of installed wind capacity but can drop down to $6\%\ -8\%$ for day-ahead wind forecasts for a single control area and to $5\%\ -7\%$ for day-ahead wind forecasts for all of Germany''\footnote{\url{http://www.nrel.gov/electricity/transmission/resource_forecasting.html}}. The accuracy may further be increased if other (commercial) tools are employed. From the previous discussion it appears reasonable to claim that on average the prediction error is below $10\%$, and this is consistent with other recent studies as well, see for instance \cite{wind} and \cite{solar}.

{\bf Comment :}  While the effect of uncertainty is beyond the scope of this paper, we note briefly that, it is simple to accommodate for forecasting errors by buying extra energy, if required, from the outer grid, or by appropriately using other storage devices, if available. However, interactions with the grid are not always convenient, either in terms of price, or in terms of environmental friendliness of the average power mix from the grid (see \cite{TITS_V2G}). An alternative to this is to formulate an uncertainty
description as part of the optimisation, and this will be part of future work.

\section{Conclusion}  \label{Conclusion}

In this paper, we introduce an optimal energy allocation scheme for the SPONGE system in the context of PHEBs. We describe a distributed AIMD algorithm for solving the optimisation problem. The main features of the proposed AIMD approach are the low-communication requirements and the privacy-preserving properties. The proposed approach is demonstrated on a case study with 15 bus trips with varying energy profiles. The results demonstrate significant environmental benefits in terms of $CO_2$ emissions that can be achieved with optimal use of free renewable energies.

\section*{Acknowledgment}

The authors gratefully acknowledge funding for this research provided by Science Foundation Ireland under grant 11/PI/1177.  The work of E. Crisostomi was supported in part by the University of Pisa under Project PRA 2016 “Sviluppo di strumenti di analisi e gestione per l’approvvigionamento e la distribuzione dell’energia.”

\ifCLASSOPTIONcaptionsoff
  \newpage
\fi



%

\bibliographystyle{ieeetran}
\bibliography{References}

%

\begin{IEEEbiography}[{\includegraphics[width=1in,height=1.25in,clip,keepaspectratio]{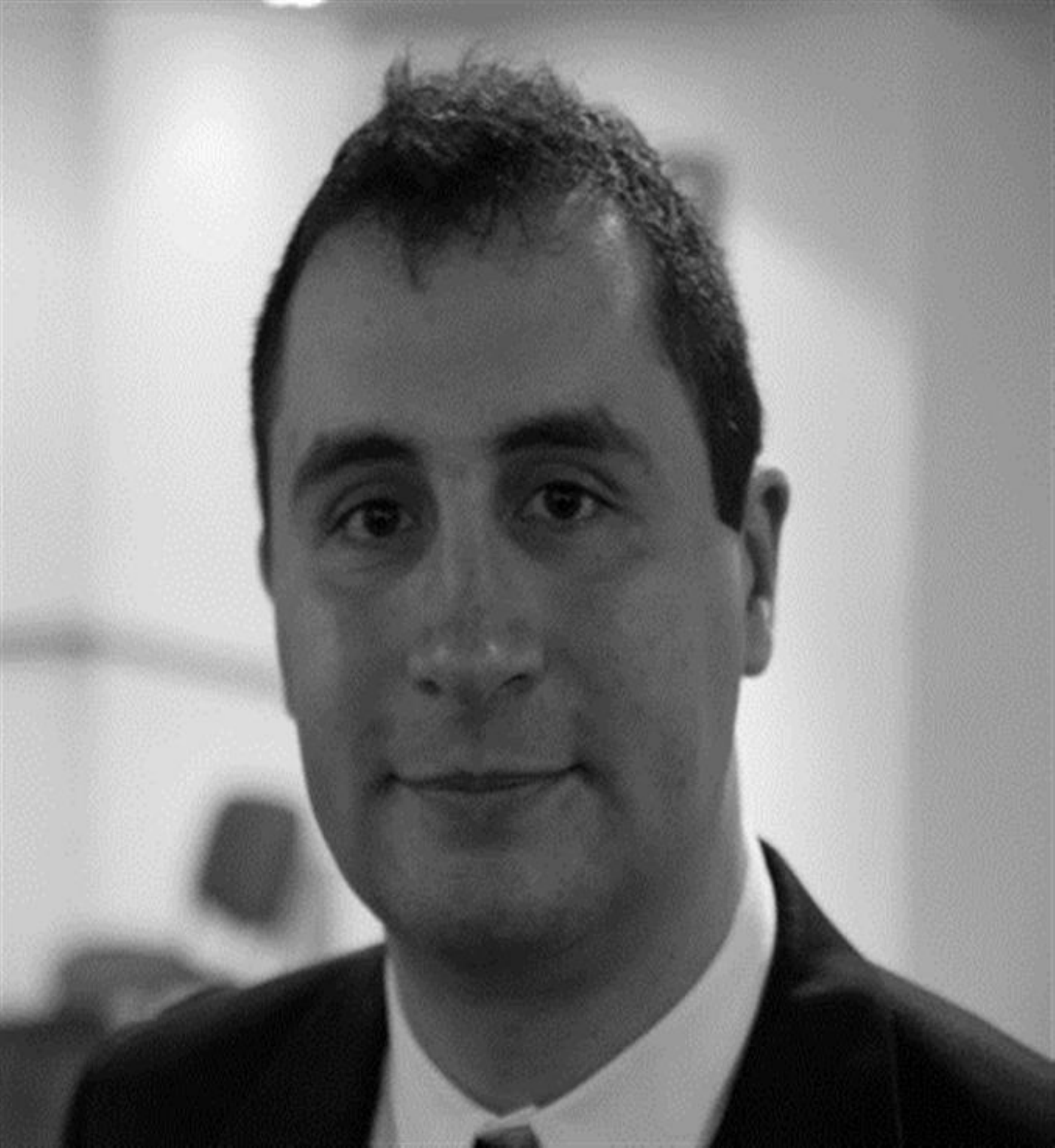}}]{Joe Naoum-Sawaya}
is currently an Assistant Professor in Management Science at Ivey Business School. He received a B.E. in Computer Engineering from the American University of Beirut and a M.A.Sc. and Ph.D. in Operations Research from the University of Waterloo. His research interests include large scale optimization methods for practical problems arising in the industry. 
\end{IEEEbiography}

\begin{IEEEbiography}[{\includegraphics[width=1in,height=1.25in,clip,keepaspectratio]{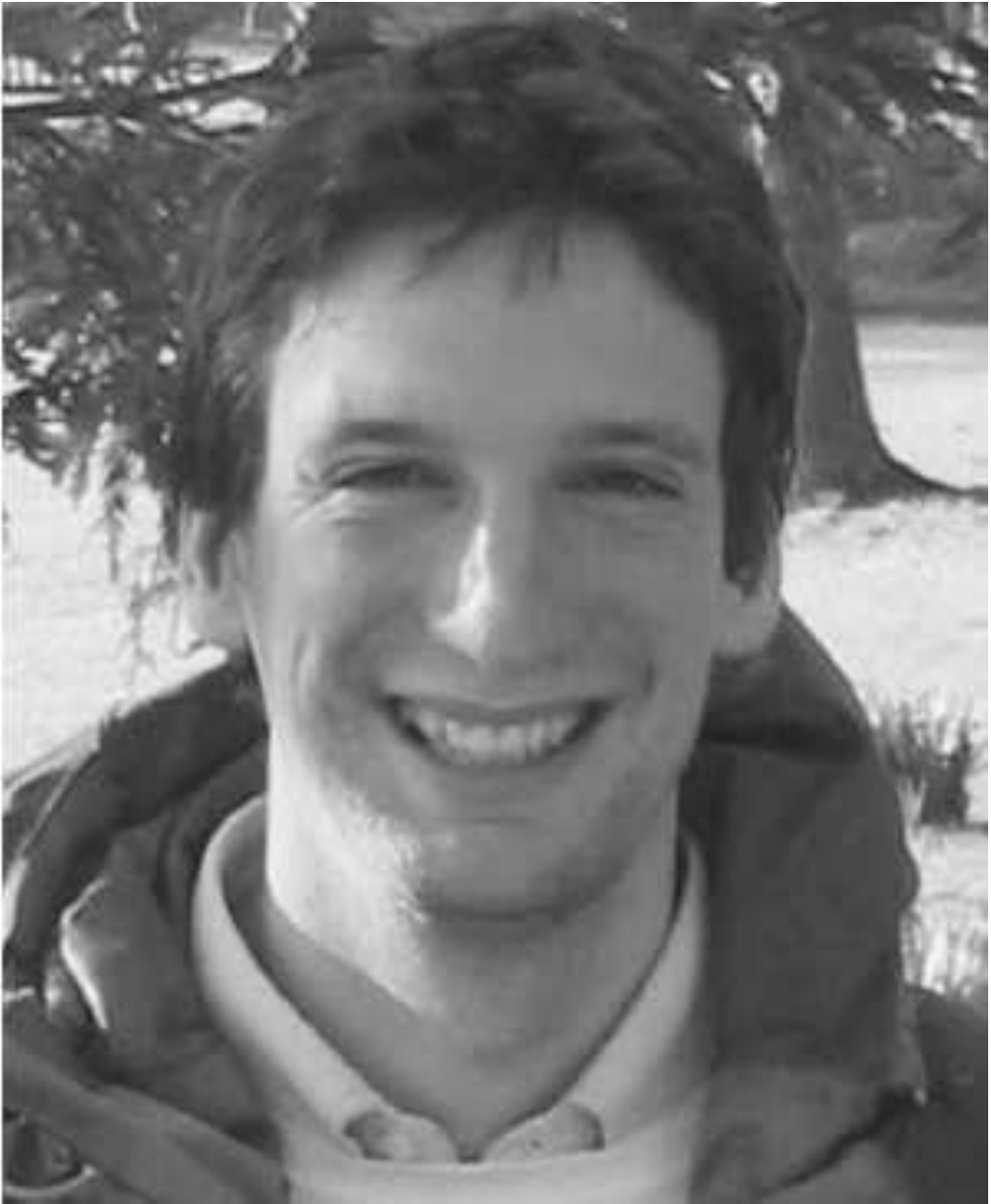}}]{Emanuele Crisostomi}
	received the B.S. degree in computer science engineering, the M.S. degree in automatic control, and the Ph.D. degree in automatics, robotics, and bioengineering, from the University of Pisa, Italy, in 2002, 2005, and 2009, respectively. He is currently an Assistant Professor of electrical engineering with the Department of Energy, Systems, Territory and Constructions Engineering, University of Pisa. His research interests include control and optimization of large-scale systems, with applications to smart grids and green mobility networks.
\end{IEEEbiography}

\begin{IEEEbiography}[{\includegraphics[width=1in,height=1.25in,clip,keepaspectratio]{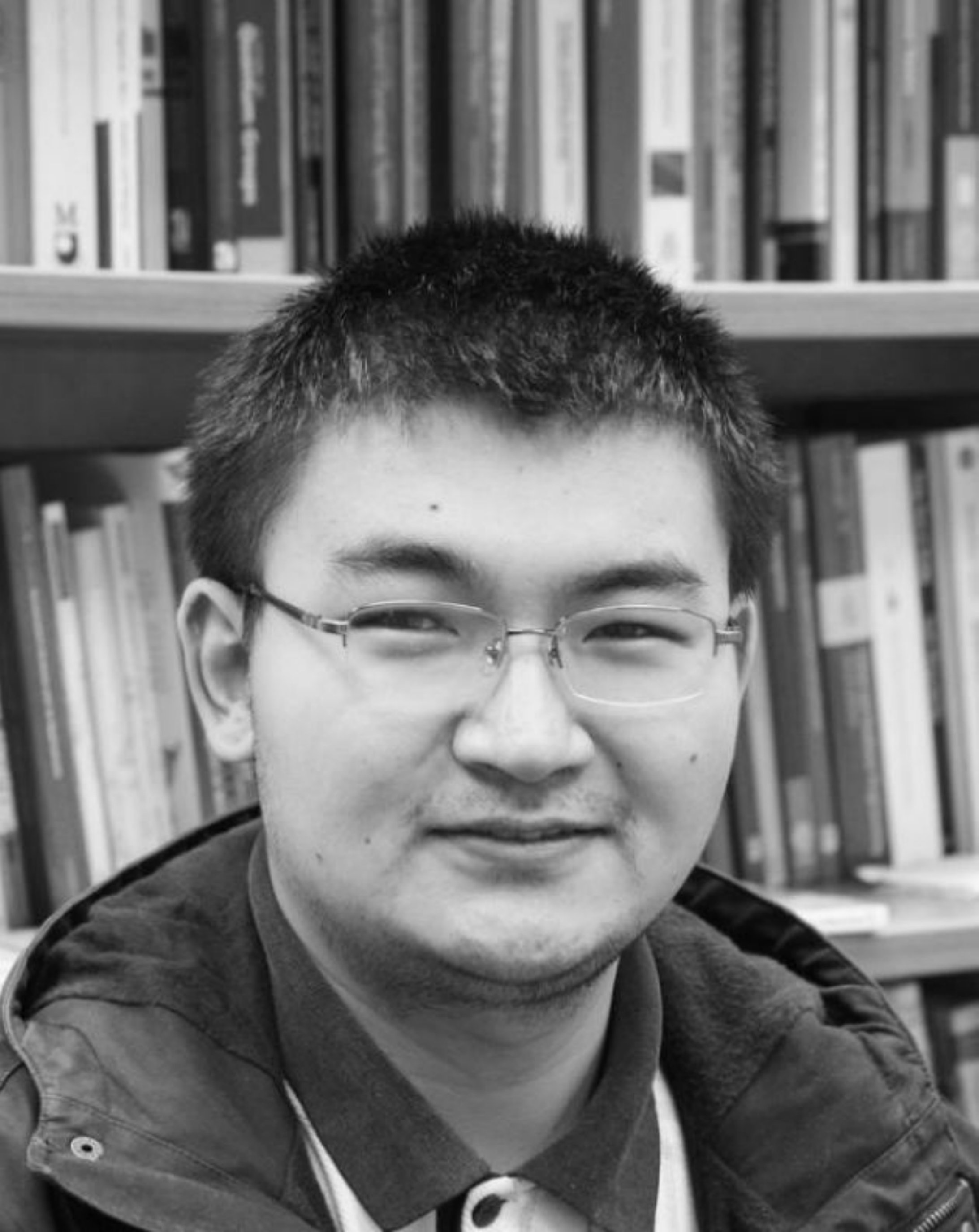}}]{Mingming~Liu}
	received his double B.E. degrees in Electronic Engineering with first class honours from National University of Ireland Maynooth and China Changzhou University in 2011. He has obtained his Ph.D. degree from the Hamilton Institute, Maynooth University. He is currently a post-doctoral researcher in University College Dublin, working with Prof. R. Shorten. His current research interests are nonlinear system dynamics, distributed control techniques, modelling and optimisation in the context of smart grid and smart transportation systems. 		
\end{IEEEbiography}

\begin{IEEEbiography}[{\includegraphics[width=1in,height=1.25in,clip,keepaspectratio]{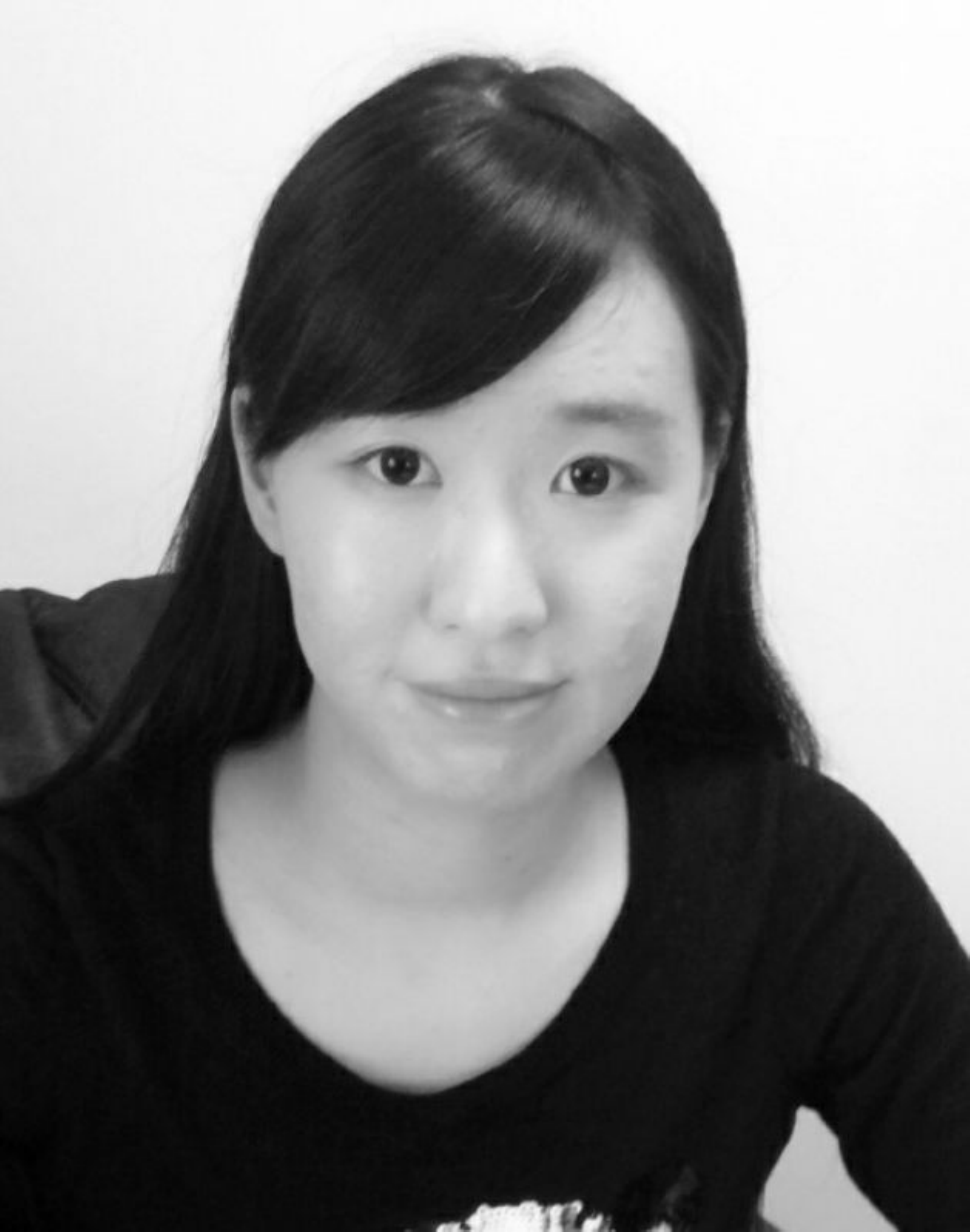}}]{Yingqi Gu}
	received her B.E. degree in Electronic Engineering (first class honours) from Maynooth University in 2013. She obtained her M.Sc degree in Signal Processing and Communications at School of Engineering, University of Edinburgh in 2014. From February 2015, she commenced her Ph.D. degree in the University College Dublin with Prof. Robert Shorten. Her current research interests are modelling, simulation and optimisation of intelligent transportation systems.   		
\end{IEEEbiography}

\begin{IEEEbiography}[{\includegraphics[width=1in,height=1.25in,clip,keepaspectratio]{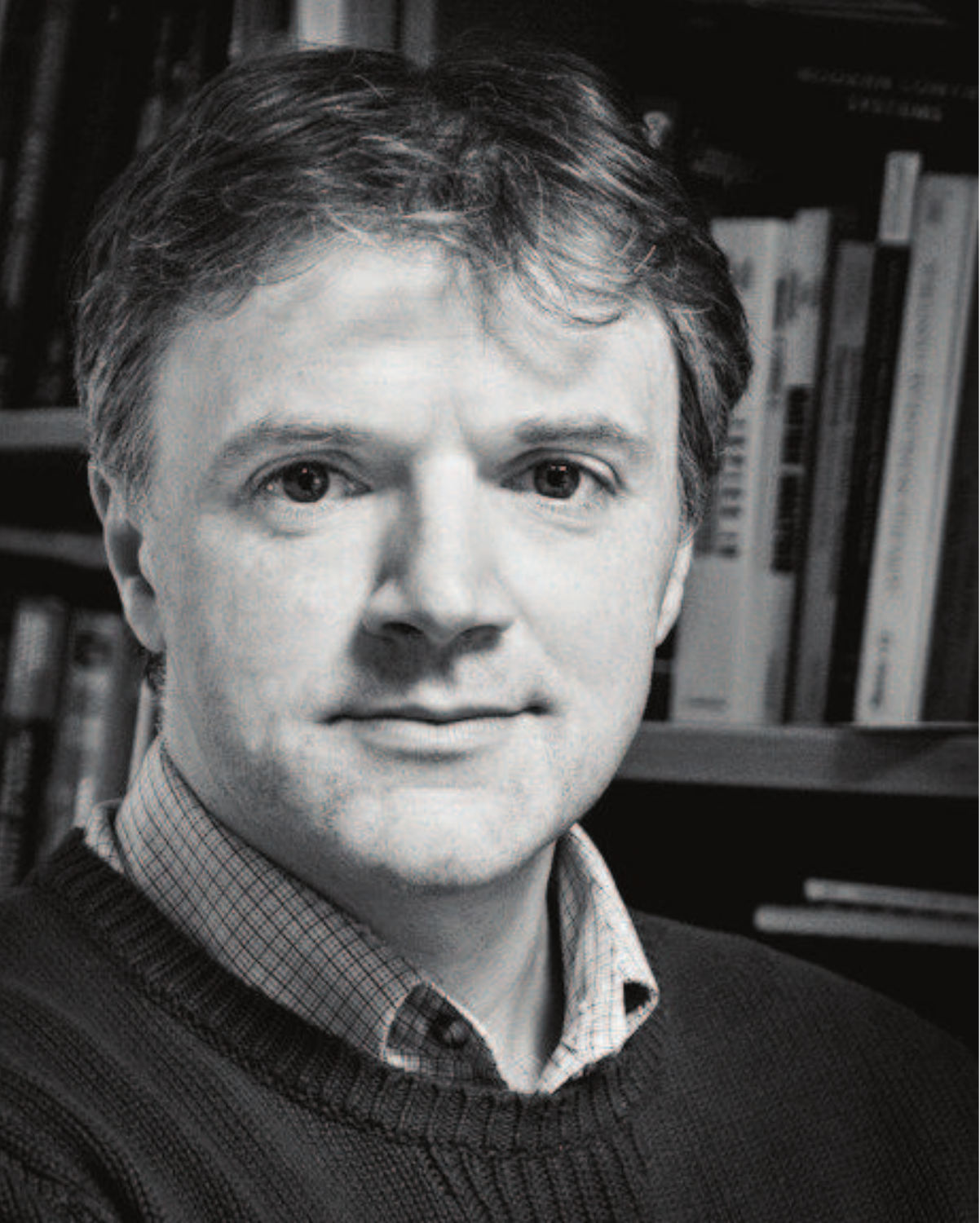}}]{Robert Shorten}
	Professor Shorten is currently Professor of Control Engineering and Decision Science at University College Dublin, and holds a position at IBM Research. Prof. Shorten's research spans a number of areas. He has been active in computer networking, automotive research, collaborative mobility (including smart transportation and electric vehicles), as well as basic control theory and linear algebra. His main field of theoretical research has been the study of hybrid dynamical systems.
\end{IEEEbiography}




\end{document}